\documentclass[a4paper,10pt]{article}
\pdfoutput=1
\usepackage{jheppub}
\usepackage{mathtools}
\usepackage{physics}
\usepackage{multirow}
\usepackage{slashed}
\usepackage{graphicx}
\usepackage{hyperref}
\usepackage{tikz}

\allowdisplaybreaks  % allows equations to break across pages

\newcommand{\orcid}[1]{%
	\href{https://orcid.org/#1}{%
		\raisebox{-0.5ex}{%
			\includegraphics[height=1.8ex]{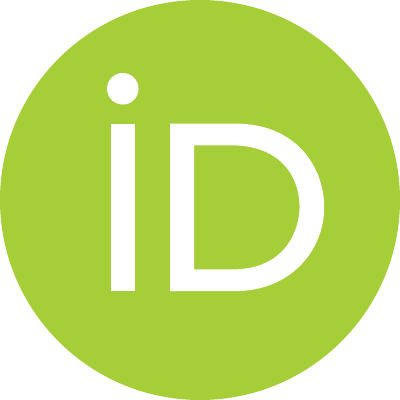}%
		}%
	}%
}

\title{\boldmath Nonleptonic $\Omega_{b}^{*}\rightarrow\Omega_{c}^{*} P(V)$ weak transitions in QCD}
\author[a]{A. Amiri\,\orcid{0000-0003-2479-4207},}
\emailAdd{amir.amiri1308@gmail.com}
\author[b,c,*]{K. Azizi\,\orcid{0000-0003-3741-2167}}
\emailAdd{kazem.azizi@ut.ac.ir}
\affiliation[a]{Department of Physics, Faculty of Science, Ferdowsi University of Mashhad, P.O.Box 1436,\\ Mashhad, Iran}
\affiliation[b]{Department of Physics, University of Tehran, North Karegar Avenue, Tehran 14395-547,\\ Iran}
\affiliation[c]{Department of Physics, Dogus University, Dudullu-$\ddot{U}$mraniye, 34775 Istanbul, T$\ddot{u}$rkiye}

\makeatletter
\newcommand*{\corrauthor}[1]{%
	\begingroup
	\renewcommand\thefootnote{*}%
	\footnotetext{Corresponding author #1}%
	\addtocounter{footnote}{-1}%
	\endgroup
}
\makeatother
\corrauthor{}

\abstract{We investigate the nonleptonic two-body weak decays of the single bottom baryon $\Omega_{b}^{*}$ into $\Omega_{c}^{*}P(V)$ final states within the factorization framework. Employing this framework and incorporating the contributions from the current-current operators, we compute the tree-level decay amplitudes and decay widths of the $\Omega_{b}^{*}\rightarrow\Omega_{c}^{*}P(V)$ processes in terms of the $\Omega_{b}^{*}\rightarrow\Omega_{c}^{*}$ transition form factors. Here, $P$ and $V$ denote pseudoscalar and vector mesons, respectively. Using the form factors obtained in our previous work, we evaluate the numerical values of the decay widths for the dominant nonleptonic weak channels. This study complements our previous analysis of the semileptonic weak transitions $\Omega_{b}^{*}\rightarrow\Omega_{c}^{*}\ell\bar{\nu}_{\ell}$ reported in Ref.~\cite{Amiri:2025gcf}, thereby providing a detailed investigation of dominant $\Omega_{b}^{*}\rightarrow\Omega_{c}^{*}$ weak decays of the $\Omega_{b}^{*}$ baryon.}
\begin{document} 
\begin{flushright}
\end{flushright}
\maketitle

%%%%%%%%%%%%%%%%%%%%%%%%%%%%%%%%%%%%%%%%%%%%%%%%%
\section{Introduction}
%%%%%%%%%%%%%%%%%%%%%%%%%%%%%%%%%%%%%%%%%%%%%%%%%
One of the central topics in heavy baryon physics is the investigation of their decay modes, particularly weak transitions, which provide essential insights into the origin of CP violation, offer stringent tests of the Standard Model (SM), constrain possible new physics effects, and assess the reliability of theoretical approaches used to describe such processes. In theoretical studies, significant progress has been achieved in understanding both the semileptonic and nonleptonic weak decay processes. For semileptonic decays, the primary challenge is the evaluation of heavy-to-light transition form factors, which encode nonperturbative QCD dynamics. These form factors have been extensively investigated using a variety of QCD-based approaches and quark models ~\cite{Amiri:2025gcf,Azizi:2009wn,Aliev:2010uy,Azizi:2018axf,Ahmadi:2025oal,Cheng:1995fe,Ebert:2006rp,Pervin:2006ie,Du:2011nj,Ke:2012wa,Han:2022srw}. Compared with semileptonic decays, the nonleptonic weak decays of heavy baryons involve more intricate QCD dynamics due to strong interactions among the final state particles. Their theoretical treatment typically relies on various factorization approaches originally developed for the study of $B$-meson decays. The study of two-body nonleptonic $B$ decays began with the naive factorization approach~\cite{Bauer:1986bm,Neubert:1997uc,Ali:1997nh}. However, this method exhibited certain limitations in describing specific decay channels and was subsequently refined through the development of QCD factorization (QCDF)~\cite{Beneke:1999br,Beneke:2000ry,Beneke:2001ev}. The factorization framework has since been successfully applied to a wide range of hadronic systems, including mesons~\cite{Beneke:1999br,AbdEl-Hady:1999jux,Du:2001hr,Beneke:2002jn,Du:2002cf,Li:2003hea,Beneke:2003zv,Ivanov:2006ni,Beneke:2006hg,Azizi:2008ty,Sun:2008wa,Khosravi:2008jw,Naimuddin:2012dy,Nayak:2022qaq}, exotic hadrons~\cite{Sundu:2019feu,Agaev:2020mqq}, and baryons~\cite{Lee:1957qs,Pakvasa:1990if,Zhu:2016bra,Zhu:2018jet,Duan:2024zjv,Li:2024htn}.

The $\Omega_{b}^{*}$ baryon is one of the fifteen single-heavy bottom baryons whose existence is predicted by the quark model; however, its definitive experimental establishment remains challenging due to current experimental limitations~\cite{Amiri:2025gcf}. Since no Okubo-Zweig-Iizuka (OZI)-allowed strong decay channels are available, this baryon is expected to decay only through electromagnetic and weak interactions. Although the radiative transition $\Omega_{b}^{*} \rightarrow \Omega_{b}\gamma$ is allowed, the small hyperfine mass splitting between the $\Omega_{b}^{*}$ and $\Omega_{b}$ states results in the emission of a very soft photon, rendering its detection extremely difficult in a hadron-collider environment~\cite{Amiri:2025gcf,Peng:2024pyl}. 

In this context, weak decays of the $\Omega_{b}^{*}$ baryon may provide a more promising avenue for its experimental identification. Unlike radiative decays, weak decays do not rely on the reconstruction of soft photons and instead produce final states containing charged hadrons and/or leptons, which can be detected with high efficiency. In particular, semileptonic decays such as $\Omega_{b}^{*}\rightarrow\Omega_{c}^{*}\ell\bar{\nu}_{\ell}$ benefit from clean lepton triggers, while nonleptonic modes of the type $\Omega_{b}^{*}\rightarrow\Omega_{c}^{*}P(V)$ allow for full reconstruction and invariant-mass analyses. Despite challenges related to limited production rates and combinatorial backgrounds, these weak decay channels may offer a viable strategy for the experimental observation of the $\Omega_{b}^{*}$ baryon, especially at experiments such as LHCb. 

Various strong, radiative, and semileptonic weak decay modes of $\Omega_{b}^{*}$ have been investigated in previous studies~\cite{Amiri:2025gcf,Peng:2024pyl,Ortiz-Pacheco:2024qcf}. In the present study, we employ the factorization approach to investigate the nonleptonic two-body weak decays of the spin-$\tfrac{3}{2}$ bottom baryon $\Omega_{b}^{*}$ to $\Omega_{c}^{*}$. This analysis is intended as a complement to our earlier study of the semileptonic weak decays $\Omega_{b}^{*}\rightarrow\Omega_{c}^{*}\ell\bar{\nu}_{\ell}$~\cite{Amiri:2025gcf}, with the aim of providing a comprehensive investigation of all possible weak decay channels of the $\Omega_{b}^{*}$ baryon. Such a theoretical analysis may contribute to and support its eventual experimental establishment. Under the diquark approximation, a baryon can be treated analogously to a meson, which allows the factorization framework to be extended naturally to heavy baryon decays. We perform a systematic study of the decays $\Omega_{b}^{*}\rightarrow\Omega_{c}^{*}P(V)$, where $P$ and $V$ denote the pseudoscalar and vector mesons, respectively. The analysis includes the contributions of the current-current operators at leading order in the effective weak Hamiltonian. Using the factorization formalism, we derive the decay amplitudes and decay widths in terms of the transition form factors of the $\Omega_{b}^{*}\rightarrow\Omega_{c}^{*}$ process. Employing the form factors previously computed in our earlier work~\cite{Amiri:2025gcf}, we then obtain numerical predictions for the decay widths of all relevant channels. In this analysis, we neglect long-distance contributions arising from interactions between the $P(V)$ meson and the $\Omega_{b}^{*}\Omega_{c}^{*}$ system, which are discussed in detail in Ref.~\cite{Duan:2024zjv}. Further information on the higher order QCD corrections in terms of $\alpha_s$ for the baryonic and mesonic decays can also be found in Refs.~\cite{Pilipp:2007mg,Bell:2008ws,Bell:2009nk,Zhu:2018jet}.

The paper is organized as follows. Section~\ref{sec:model} presents the theoretical framework of this study, including the effective weak Hamiltonian, the  factorization approach, and the derivation of the decay amplitudes and decay widths. In Section~\ref{sec:Num}, we provide the numerical analysis, where the form factors, decay widths are evaluated. The main conclusions are summarized in Section~\ref{sec:conclusions}. Finally, Appendix~\ref{Appen} contains the explicit expressions for the squared decay amplitudes.
%%%%%%%%%%%%%%%%%%%%%%%%%%%%%%%%%%%%%%%%%%%%%%%%%
\section{Theoretical framework}\label{sec:model}
%%%%%%%%%%%%%%%%%%%%%%%%%%%%%%%%%%%%%%%%%%%%%%%%%
In this section, we outline the theoretical framework of our analysis, including a brief review of the effective weak Hamiltonian, the factorization approach, and the derivation of the decay amplitudes and decay widths for the nonleptonic weak decays of the $\Omega_{b}^{*}$ baryon.
%%%%%%%%%%%%%%%%%%%%%%%%%%%%%%%%%%%%%%%%%%%
\subsection{Nonleptonic decays of $\Omega_{b}^{*}$}
%%%%%%%%%%%%%%%%%%%%%%%%%%%%%%%%%%%%%%%%%%%
The dominant nonleptonic weak decays of $\Omega_{b}^{*}$ proceed via the underlying quark-level transition $b\rightarrow W^- c$. Accordingly, in this study we focus on the nonleptonic decays $\Omega_{b}^{*}\rightarrow\Omega_{c}^{*}P(V)$. In these processes, $P$ denotes the pseudoscalar mesons $\pi^-$, $K^-$, $D^-$ and $D_s^-$, while $V$ represents the vector mesons $\rho^-$, $K^{*-}$, $D^{*-}$ and $D_s^{*-}$. These pseudoscalar and vector mesons in the final state originate from the hadronization of the $W^-$ boson, which decays into quark-antiquark pairs $d\bar{u}$, $s\bar{u}$, $d\bar{c}$ and $s\bar{c}$, respectively. 
 
%%%%%%%%%%%%%%%%%%%%%%%%%%%%%%%%%%%%%%%%%%%
\subsection{Effective weak Hamiltonian}\label{sec:EWH}
%%%%%%%%%%%%%%%%%%%%%%%%%%%%%%%%%%%%%%%%%%%
To describe the weak decays of the $\Omega_{b}^{*}$ baryon, it is essential to consider the corresponding effective weak Hamiltonian. In these processes, three distinct energy scales are involved: $m_W\gg m_b \gg \Lambda_{QCD}$. The most suitable framework for such a multiscale problem is the Effective Field Theory (EFT) approach, where the high energy degrees of freedom are integrated out and the interactions are expressed through a series of local effective operators using the Operator Product Expansion (OPE). In this formalism, all short-distance (high energy) effects above the scale $m_b$ are encapsulated in the Wilson coefficients, which can be computed perturbatively order by order through matching at the quark level. Accordingly, the effective weak Hamiltonian governing the nonleptonic weak decays of $\Omega_{b}^{*}$, corresponding to the underlying $b\rightarrow c$ transition at the tree-level, is given by~\cite{Buchalla:1995vs}, 
\begin{equation}
	\label{EffH}
	\mathcal{H}_{eff}  = \frac{G_F}{\sqrt{2}} V_{bc} V_{qq'}^* \big[C_1(\mu) Q_1 + C_2(\mu) Q_2\big]  \,.
\end{equation}
The coefficients $C_1(\mu)$ and $C_2(\mu)$ are the Wilson coefficients evaluated at the renormalization scale $\mu$, while the corresponding current-current operators, $Q_1$ and $Q_2$ are defined as,
\begin{equation}
	\label{ccope}
	Q_1  = (\bar{q}_i q'_i)_{V-A} (\bar{c}_j b_j)_{V-A},\ \ \ Q_2  = (\bar{q}_i q'_j)_{V-A} (\bar{c}_j b_i)_{V-A} \,,
\end{equation} 
where $q = d,s$ and $q' = u,c$, while $i$, $j$ denote the color indices. The vector minus axial-vector current is given by $(\bar{q}_{1} q_{2})_{V-A} =\bar{q}_{1} \gamma_{\mu}(1-\gamma_5) q_{2}$.

The theoretical description of these tree-level nonleptonic weak decays of the $\Omega_{b}^{*}$ baryon can be formulated through the evaluation of the matrix elements of the local effective operators between appropriate hadronic states. The amplitudes of such processes can be effectively computed using the naive factorization approach, in which the decay amplitude is expressed as the product of the decay constant of the pseudoscalar (or vector) mesons $P (V)$ and the weak transition form factors governing the $\Omega_{b}^{*}\rightarrow\Omega_{c}^{*}$ transition. 
 
%%%%%%%%%%%%%%%%%%%%%%%%%%%%%%%%%%%%%%%%%%%
\subsection{Factorization approach}\label{sec:QCDF}
%%%%%%%%%%%%%%%%%%%%%%%%%%%%%%%%%%%%%%%%%%%
In nonleptonic weak decays of hadrons, the final state typically involves at least three hadrons. Consequently, the evaluation of the hadronic matrix elements of the local operators appearing in the effective Hamiltonian, Eq.~(\ref{EffH}), represents a nonperturbative and technically challenging problem. To simplify these challenging calculations, the factorization approach was proposed. The earliest and most straightforward version of this method is the naive factorization approach~\cite{Bauer:1986bm,Neubert:1997uc,Ali:1997nh}. To illustrate the underlying idea, consider a typical two-body decay of a heavy meson, such as $M\rightarrow M_1M_2$, where $M_1$ denotes the recoiling meson containing the spectator quark, and $M_2$ represents the emitted meson produced directly from the weak current. The key assumption of factorization is that the emitted meson $M_2$ decouples from the remaining $MM_1$ system, a simplification known as the vacuum insertion approximation. Under this assumption, the three hadron matrix element factorizes into the product of a transition form factor and a decay constant. Although the naive factorization approach describes color-allowed tree-level processes reasonably well ~\cite{Bauer:1986bm,Neubert:1997uc,Ali:1997nh,Khosravi:2008jw}, it fails for color-suppressed and penguin-induced transitions~\cite{Zhu:2018jet}. 

In this study, both the initial and final states involve heavy baryons, while the emitted meson carries large momentum in the heavy-quark limit. Under the diquark approximation, a heavy baryon can be treated analogously to a heavy meson, which allows the extension of factorization arguments to baryonic decays. In this framework, soft gluon interactions between the energetic emitted meson and the recoiling baryonic system are suppressed due to color transparency, particularly for energetic pseudoscalar or vector mesons. Consequently, the leading contribution to the decay amplitude factorizes into the product of a meson decay constant and a baryonic transition matrix element. The inclusion of hard spectator-scattering effects would require modeling the internal structure of the diquark via an additional form factor, which is currently not well constrained from first principles. Therefore, following established studies of nonleptonic heavy baryon decays, such contributions are commonly neglected at leading order~\cite{Zhu:2018jet,Duan:2024zjv}. 

For the tree-level, color-allowed processes induced solely by current-current operators in the $\Omega_{b}^{*}\rightarrow\Omega_{c}^{*}P(V)$ transitions considered here, we adopt the naive factorization approach to compute the nonleptonic decay amplitudes. It is well known that, for such modes,
naive factorization provides a reasonable first approximation, as nonfactorizable contributions
are expected to be suppressed.
%%%%%%%%%%%%%%%%%%%%%%%%%%%%%%%%%%%%%%%%%%%
\subsection{Decay amplitudes and decay widths}
%%%%%%%%%%%%%%%%%%%%%%%%%%%%%%%%%%%%%%%%%%%
The decay amplitudes of the weak transition $\Omega_{b}^{*}\rightarrow\Omega_{c}^{*}P(V)$, corresponding to each four-quark operator in the effective Hamiltonian given in Eq.~(\ref{EffH}), can be generally expressed in terms of the hadronic matrix elements of these operators as follows,
\begin{equation}
	\label{Ampge}
	\mathcal{A}_{l}(\Omega_{b}^{*}\rightarrow\Omega_{c}^{*}P(V)) = \langle \Omega_{c}^{*} P(V) | \mathcal{H}_{eff} | \Omega_{b}^{*} \rangle = \frac{G_F}{\sqrt{2}} \ V_{CKM} \sum_{l} C_{l} \langle \Omega_{c}^{*} P(V) | Q_{l} | \Omega_{b}^{*} \rangle  \,.
\end{equation} 
As discussed in subsections~\ref{sec:EWH} and~\ref{sec:QCDF}, within the naive factorization approximation, these hadronic matrix elements can be decomposed into the product of the decay constant of the emitted meson $P (V)$ and the form factors describing the $\Omega_{b}^{*}\rightarrow\Omega_{c}^{*}$ transition,
\begin{equation}
	\label{MatE}
	\langle \Omega_{c}^{*} P(V) | Q_{l} | \Omega_{b}^{*} \rangle = \langle P(V) | (\bar{q}_{r} q'_{s})_{V-A} | 0 \rangle \times \langle \Omega_{c}^{*} | (\bar{c}_{r'} b_{s'})_{V-A} | \Omega_{b}^{*} \rangle \,.
\end{equation} 
By evaluating the hadronic matrix elements $\langle \Omega_{c}^{*} P(V) | Q_{l} | \Omega_{b}^{*} \rangle$ in Eq.~(\ref{Ampge}) for each current-current operator $Q_{1,2}$ using Eq.~(\ref{MatE}) and performing a Fierz transformation to match the flavor quantum numbers of the currents with those of the physical hadrons, the decay amplitude for the $\Omega_{b}^{*}\rightarrow\Omega_{c}^{*}P(V)$ transition can be written as:
\begin{align} \label{DecAmp} 
	\mathcal{A}&(\Omega_{b}^{*}\rightarrow\Omega_{c}^{*}P(V))= \frac{G_F}{\sqrt{2}} V_{bc} V_{qq'}^*  \big( \langle \Omega_{c}^{*} P(V) | C_1(\mu) Q_1 | \Omega_{b}^{*} \rangle +\langle \Omega_{c}^{*} P(V) | C_2(\mu) Q_2 | \Omega_{b}^{*} \rangle \big) \nonumber \\& = \frac{G_F}{\sqrt{2}} V_{bc} V_{qq'}^* a_1(\mu)\ \langle P(V) | (\bar{q}_i q'_i)_{V-A} | 0 \rangle \times \langle \Omega_{c}^{*} | (\bar{c}_j b_j)_{V-A} | \Omega_{b}^{*} \rangle   \,.
\end{align}	
For pseudoscalar and vector mesons, the decay amplitudes are explicitly given by,
\begin{equation}
	\label{DecAmpPsem}
	\begin{split}
		\mathcal{A}(\Omega_{b}^{*}(p)\rightarrow\Omega_{c}^{*}(p')P(q)) =& \frac{G_F}{\sqrt{2}} V_{bc} V_{qq'}^* a_1(\mu)\ \langle P(q) | \bar{q}_{i} \gamma_{\mu}(1-\gamma_5) q'_{i} | 0 \rangle \\& \times \langle \Omega_{c}^{*}(p') | \bar{c}_{j} \gamma_{\mu}(1-\gamma_5) b_{j} | \Omega_{b}^{*}(p) \rangle   \,,
	\end{split}
\end{equation}
\begin{equation}
	\label{DecAmpVecm}
	\begin{split}
		\mathcal{A}(\Omega_{b}^{*}(p)\rightarrow\Omega_{c}^{*}(p')V(q)) =& \frac{G_F}{\sqrt{2}} V_{bc} V_{qq'}^* a_1(\mu)\ \langle V(q) | \bar{q}_{i} \gamma_{\mu}(1-\gamma_5) q'_{i} | 0 \rangle \\& \times \langle \Omega_{c}^{*}(p') | \bar{c}_{j} \gamma_{\mu}(1-\gamma_5) b_{j} | \Omega_{b}^{*}(p) \rangle \,,
	\end{split}
\end{equation}
where $a_1(\mu)$ denotes the effective Wilson coefficient combination associated with the color-allowed tree-level contribution, defined as,
\begin{equation}\label{Coea}
	a_1(\mu) = C_1(\mu) + \frac{1}{N_c} C_2(\mu) \,,
\end{equation}
with $N_c = 3$ being the number of quark colors. In Eqs.~(\ref{DecAmpPsem}) and~(\ref{DecAmpVecm}), the first matrix element can be parametrized in terms of the mesons decay constants as follows,
\begin{equation}\label{PseME}
	\langle P(q) | \bar{q}_{i} \gamma_{\mu}(1-\gamma_5) q'_{i} | 0 \rangle = i f_P q_{\mu} \,,
\end{equation}
\begin{equation}\label{VeME}
	\langle V(q) | \bar{q}_{i} \gamma_{\mu}(1-\gamma_5) q'_{i} | 0 \rangle = m_V f_V \epsilon_{\mu}^{*} \,,
\end{equation}
where $f_P$ and $f_V$ denote the decay constant of the pseudoscalar and vector mesons, respectively, while $m_V$ and $\epsilon_{\mu}$ represent the mass and polarization vector of the vector meson. The second matrix element, $\langle \Omega_{c}^{*}(p') | \bar{c}_{j} \gamma_{\mu}(1-\gamma_5) b_{j} | \Omega_{b}^{*}(p) \rangle$, encodes the nonperturbative dynamics of the $\Omega_{b}^{*}\rightarrow\Omega_{c}^{*}$ weak transition and is expressed in terms of the corresponding form factors~\cite{Amiri:2025gcf}, 
\begin{align} \label{BBME}
	  &\langle \Omega_{c}^{*}(p') | \bar{c}_{j} \gamma_{\mu}(1-\gamma_5) b_{j} | \Omega_{b}^{*}(p) \rangle = \bar{u}_{\Omega_{c}^{*}}^{\alpha}(p',s')\ \Big[g_{\alpha\beta}\Big(\gamma_{\mu}F_{1}(q^2)-i\sigma_{\mu\nu}\frac{q_\nu}{m_{\Omega_{b}^{*}}}F_{2}(q^2)+ \frac{q_\mu}{m_{\Omega_{b}^{*}}}F_{3}(q^2)\Big) \nonumber \\& + \frac{q_{\alpha}q_{\beta}}{m_{\Omega_{b}^{*}}^2}\ \Big(\gamma_{\mu}F_{4}(q^2)-i\sigma_{\mu\nu}\frac{q_\nu}{m_{\Omega_{b}^{*}}}F_{5}(q^2)+ \frac{q_\mu}{m_{\Omega_{b}^{*}}}F_{6}(q^2)\Big) + \frac{(g_{\alpha\mu}q_{\beta} - g_{\beta\mu} q_{\alpha})} {m_{\Omega_{b}^{*}}} F_{7}(q^2)\Big]\ u_{\Omega_{b}^{*}}^{\beta}(p,s) \nonumber \\& - \bar{u}_{\Omega_{c}^{*}}^{\alpha}(p',s')\ \Big[g_{\alpha\beta}\Big(\gamma_{\mu}G_{1}(q^2)-i\sigma_{\mu\nu}\frac{q_\nu}{m_{\Omega_{b}^{*}}}G_{2}(q^2)+ \frac{q_\mu}{m_{\Omega_{b}^{*}}}G_{3}(q^2)\Big) + \frac{q_{\alpha}q_{\beta}}{m_{\Omega_{b}^{*}}^2}\ \Big(\gamma_{\mu}G_{4}(q^2) \nonumber \\& -i\sigma_{\mu\nu}\frac{q_\nu}{m_{\Omega_{b}^{*}}}G_{5}(q^2)+ \frac{q_\mu}{m_{\Omega_{b}^{*}}}G_{6}(q^2)\Big) + \frac{(g_{\alpha\mu}q_{\beta} - g_{\beta\mu} q_{\alpha})} {m_{\Omega_{b}^{*}}} G_{7}(q^2)\Big]\ \gamma_5\ u_{\Omega_{b}^{*}}^{\beta}(p,s) \,,
\end{align}	
where $q = p - p'$ denotes the transferred momentum, with $p$ and $p'$ being the four-momenta of the initial and final baryons, respectively.

At this stage, the decay amplitudes for the $\Omega_{b}^{*}\rightarrow\Omega_{c}^{*}P(V)$ transitions can be computed using the relations introduced above. For the weak decay $\Omega_{b}^{*}\rightarrow\Omega_{c}^{*}P$, where the final state contains a pseudoscalar meson, substituting Eqs.~(\ref{PseME}) and~(\ref{BBME}) into Eq.~(\ref{DecAmpPsem}) and performing the required algebraic manipulations lead to the following expression for the decay amplitude:
\begin{align}	\label{FDecAmpPsem}
		\mathcal{A}_P&\big(\Omega_{b}^{*}(p)\rightarrow\Omega_{c}^{*}(p')P(q)\big) = i \frac{G_F}{\sqrt{2}} V_{bc} V_{qq'}^* \ a_1(\mu) f_P \Big[\Big((m_{\Omega_{b}^{*}}-m_{\Omega_{c}^{*}})F_1(q^2)+\frac{q^2}{m_{\Omega_{b}^{*}}} F_3(q^2)\Big) \nonumber \\& g_{\alpha\beta}\bar{u}_{\Omega_{c}^{*}}^{\alpha}(p',s')u_{\Omega_{b}^{*}}^{\beta}(p,s) - \Big(\frac{m_{\Omega_{b}^{*}}-m_{\Omega_{c}^{*}}}{m_{\Omega_{b}^{*}}^2}F_4(q^2) + \frac{q^2}{m_{\Omega_{b}^{*}}^3}F_6(q^2)\Big)p_{\alpha}p'_{\beta}\bar{u}_{\Omega_{c}^{*}}^{\alpha}(p',s')u_{\Omega_{b}^{*}}^{\beta}(p,s) \nonumber \\& + \Big((m_{\Omega_{b}^{*}}+m_{\Omega_{c}^{*}})G_1(q^2)-\frac{q^2}{m_{\Omega_{b}^{*}}} G_3(q^2)\Big)  g_{\alpha\beta}\bar{u}_{\Omega_{c}^{*}}^{\alpha}(p',s') \gamma_5 u_{\Omega_{b}^{*}}^{\beta}(p,s) \nonumber \\& - \Big(\frac{m_{\Omega_{b}^{*}}+m_{\Omega_{c}^{*}}}{m_{\Omega_{b}^{*}}^2}G_4(q^2) - \frac{q^2}{m_{\Omega_{b}^{*}}^3}G_6(q^2)\Big)p_{\alpha}p'_{\beta}\bar{u}_{\Omega_{c}^{*}}^{\alpha}(p',s') \gamma_5 u_{\Omega_{b}^{*}}^{\beta}(p,s)\Big] \,.
\end{align}
For the weak decay $\Omega_{b}^{*}\rightarrow\Omega_{c}^{*}V$, where the final state contains a vector meson, substituting Eqs.~(\ref{VeME}) and~(\ref{BBME}) into Eq.~(\ref{DecAmpVecm}) and carrying out the necessary algebraic manipulations yield:
\begin{align}	\label{FDecAmpVecm}
		&\mathcal{A}_V\big(\Omega_{b}^{*}(p)\rightarrow\Omega_{c}^{*}(p')V(q)\big) = \frac{G_F}{\sqrt{2}} V_{bc} V_{qq'}^* \ a_1(\mu) m_V f_V \epsilon^{*\mu} \Big[\Big(F_1(q^2)+\frac{m_{\Omega_{b}^{*}}+m_{\Omega_{c}^{*}}}{m_{\Omega_{b}^{*}}} F_2(q^2)\Big)\nonumber \\& g_{\alpha\beta}\bar{u}_{\Omega_{c}^{*}}^{\alpha}(p',s') \gamma_{\mu} u_{\Omega_{b}^{*}}^{\beta}(p,s) - \frac{2}{m_{\Omega_{b}^{*}}} F_2(q^2) p'_{\mu}  g_{\alpha\beta}\bar{u}_{\Omega_{c}^{*}}^{\alpha}(p',s')u_{\Omega_{b}^{*}}^{\beta}(p,s)\nonumber \\&- \Big(\frac{1}{m_{\Omega_{b}^{*}}^2}F_4(q^2) + \frac{m_{\Omega_{b}^{*}}+m_{\Omega_{c}^{*}}}{m_{\Omega_{b}^{*}}^3}F_5(q^2)\Big)p_{\alpha}p'_{\beta}\bar{u}_{\Omega_{c}^{*}}^{\alpha}(p',s') \gamma_{\mu} u_{\Omega_{b}^{*}}^{\beta}(p,s)\nonumber \\& + \frac{2}{m_{\Omega_{b}^{*}}^3} F_5(q^2) p'_{\mu}p_{\alpha}p'_{\beta}\bar{u}_{\Omega_{c}^{*}}^{\alpha}(p',s') u_{\Omega_{b}^{*}}^{\beta}(p,s) - \frac{1}{m_{\Omega_{b}^{*}}} \big(g_{\alpha\mu}p'_{\beta} + g_{\beta\mu} p_{\alpha}\big) F_7(q^2) \bar{u}_{\Omega_{c}^{*}}^{\alpha}(p',s') u_{\Omega_{b}^{*}}^{\beta}(p,s)\nonumber \\&  + \Big(-G_1(q^2)+\frac{m_{\Omega_{b}^{*}}-m_{\Omega_{c}^{*}}}{m_{\Omega_{b}^{*}}} G_2(q^2)\Big)  g_{\alpha\beta}\bar{u}_{\Omega_{c}^{*}}^{\alpha}(p',s') \gamma_{\mu} \gamma_5 u_{\Omega_{b}^{*}}^{\beta}(p,s)\nonumber \\& + \frac{2}{m_{\Omega_{b}^{*}}} G_2(q^2) p'_{\mu}  g_{\alpha\beta}\bar{u}_{\Omega_{c}^{*}}^{\alpha}(p',s') \gamma_5 u_{\Omega_{b}^{*}}^{\beta}(p,s)\nonumber \\& + \Big(\frac{1}{m_{\Omega_{b}^{*}}^2}G_4(q^2) - \frac{m_{\Omega_{b}^{*}}-m_{\Omega_{c}^{*}}}{m_{\Omega_{b}^{*}}^3}G_5(q^2)\Big)p_{\alpha}p'_{\beta}\bar{u}_{\Omega_{c}^{*}}^{\alpha}(p',s') \gamma_{\mu} \gamma_5 u_{\Omega_{b}^{*}}^{\beta}(p,s)\nonumber \\& - \frac{2}{m_{\Omega_{b}^{*}}^3} G_5(q^2) p'_{\mu}p_{\alpha}p'_{\beta}\bar{u}_{\Omega_{c}^{*}}^{\alpha}(p',s') \gamma_5 u_{\Omega_{b}^{*}}^{\beta}(p,s)  + \frac{1}{m_{\Omega_{b}^{*}}} \big(g_{\alpha\mu}p'_{\beta} + g_{\beta\mu} p_{\alpha}\big) G_7(q^2) \bar{u}_{\Omega_{c}^{*}}^{\alpha}(p',s') \gamma_5 u_{\Omega_{b}^{*}}^{\beta}(p,s)\Big]    \,.
\end{align}

The decay widths of the nonleptonic $\Omega_{b}^{*}\rightarrow\Omega_{c}^{*}P(V)$ transitions can be obtained by applying the following general relations:
\begin{equation}
	\label{DeWidPsem}
	\Gamma(\Omega_{b}^{*}\rightarrow\Omega_{c}^{*}P) = \frac{1}{64 \pi m_{\Omega_{b}^{*}}^{3}} |\mathcal{A}_P|^2 \lambda^{\frac{1}{2}}(m_{\Omega_{b}^{*}}^2,m_{\Omega_{c}^{*}}^{2},m_{P}^{2}) \,,
\end{equation}
\begin{equation}
	\label{DeWidVecm}
	\Gamma(\Omega_{b}^{*}\rightarrow\Omega_{c}^{*}V) = \frac{1}{64 \pi m_{\Omega_{b}^{*}}^{3}} |\mathcal{A}_V|^2 \lambda^{\frac{1}{2}}(m_{\Omega_{b}^{*}}^2,m_{\Omega_{c}^{*}}^{2},m_{V}^{2}) \,,
\end{equation}
where $\lambda (x,y,z) = x^{2}+y^{2}+z^{2}-2(xy+xz+yz)$ is the usual triangle function.
We evaluate Eqs.~(\ref{FDecAmpPsem}) and~(\ref{DeWidPsem}) at $q^2= m_P^2$ for pseudoscalar mesons, and Eqs.~(\ref{FDecAmpVecm}),~(\ref{DeWidVecm}) at $q^2= m_V^2$ for vector mesons, corresponding to the physical kinematics of the respective final states. The explicit expressions for the squared amplitudes, $|\mathcal{A}_P|^2$ and $|\mathcal{A}_V|^2$, are provided in Appendix~\ref{Appen}. 
%%%%%%%%%%%%%%%%%%%%%%%%%%%%%%%%%%%%%%%%%%%%%%%%%
\section{Numerical Results}\label{sec:Num}
%%%%%%%%%%%%%%%%%%%%%%%%%%%%%%%%%%%%%%%%%%%%%%%%%
In this section, we first summarize the input parameters employed in our numerical analysis. Utilizing these parameters, we calculate the decay widths for the two categories of nonleptonic $\Omega_{b}^{*}$ decays, namely $\Omega_{b}^{*}\rightarrow\Omega_{c}^{*}P(V)$, by applying Eqs.~(\ref{eq:SquAmPseM}),~(\ref{DeWidPsem}),~(\ref{eq:SquAmVecM}), and~(\ref{DeWidVecm}).
%%%%%%%%%%%%%%%%%%%%%%%%%%%%%%%%%%%%%%%%%%%
\subsection{Input parameters}
%%%%%%%%%%%%%%%%%%%%%%%%%%%%%%%%%%%%%%%%%%%
The input parameters employed in the numerical calculations are summarized in Table~\ref{tab:MesonPar1} ~\cite{Agaev:2020mqq,Duan:2024zjv,ParticleDataGroup:2024cfk}. This table lists the decay constants and masses of the final state mesons, as well as the relevant CKM matrix elements. 
%%%%%%%%%%%%%%%%%%%%%%%%%%%%%%%%%%%%%%%%%%%
\begin{table}[tbp]
	\centering
	\begin{tabular}{|c|c|c|c|c|}
		\hline\hline
		meson& $f_{P(V)}(\mathrm{MeV})$ & $m_{P(V)}(\mathrm{MeV})$ & Quantity & Value  \\ \hline\hline
		$\pi^-$ & $131$ & $139.57\pm0.00017$ & $m_{\Omega_{b}^{*}}$ & $(6084\pm84)\ $MeV \\
		$K^-$ & $155.72\pm0.51$ & $493.677\pm0.013$ & $m_{\Omega_{c}^{*}}$ & $(2765.9\pm2)\ $MeV \\
		$D^-$ & $203.7\pm4.7$ &$1869.66\pm0.05$ & $G_F$& $1.17 \times 10^{-5}\ $ $\mathrm{GeV}^{-2}$ \\
		$D_s^-$ & $257.8\pm 4.1$ &$1968.30\pm0.07$ & $|V_{bc}|$& $0.0422\pm0.00008$ \\
		$\rho^-$ & $216$  & $775.26\pm0.23$ & $|V_{ud}|$ & $0.9742\pm0.00021$\\
		$K^{*-}$ & $210$ & $891.67\pm0.26$ & $|V_{us}|$ & $0.2243\pm0.0005$\\
		$D^{*-}$ & $230$ & $2010.26\pm0.05$ &  $|V_{cd}|$ & $0.218\pm0.004$\\
		$D_s^{*-}$ & $271$ & $2112.2\pm0.4$ & $|V_{cs}|$& $0.997\pm0.017$\\
        \hline\hline
	\end{tabular}%
	\caption{Decay constants  and masses of the final state pseudoscalar and vector mesons. The CKM matrix elements are also included. }
	\label{tab:MesonPar1}
\end{table}
%%%%%%%%%%%%%%%%%%%%%%%%%%%%%%%%%%%%%%%%%%%
The Wilson coefficients $C_1(\mu=m_b)$ and $C_2(\mu=m_b)$, including next-to-leading order QCD corrections, are taken from Ref.~\cite{Agaev:2020mqq},
\begin{equation}
	C_1(m_b) = 1.117,\ \ \ \ C_2(m_b) = -0.257 \,.
\end{equation}
\subsection{Form factors}
%%%%%%%%%%%%%%%%%%%%%%%%%%%%%%%%%%%%%%%%%%%
An essential ingredient in the calculation of the decay widths for the $\Omega_{b}^{*}\rightarrow\Omega_{c}^{*}P(V)$ transitions is the set of form factors governing the $\Omega_{b}^{*}\rightarrow\Omega_{c}^{*}$ transition. As introduced in Eq.~(\ref{BBME}), these form factors capture the nonperturbative dynamics of the baryonic transition. In the present work, we employ the form factors obtained in our previous study~\cite{Amiri:2025gcf}, where they were calculated in detail for the semileptonic $\Omega_{b}^{*}\rightarrow\Omega_{c}^{*}\ell\bar{\nu}_{\ell}$ decays. These form factors are parameterized as follows,
\begin{equation}
	\mathcal{F}_{l}(q^2) = \frac{\mathcal{F}_{l}(0)}{1-a\ \Big(\frac{q^2}{m_{\Omega_{b}^{*}}^{2}}\Big) +b\ \Big(\frac{q^2}{m_{\Omega_{b}^{*}}^{2}}\Big)^2+c\ \Big(\frac{q^2}{m_{\Omega_{b}^{*}}^{2}}\Big)^3+d\ \Big(\frac{q^2}{m_{\Omega_{b}^{*}}^{2}}\Big)^4} \,. 
\end{equation}
The parameters $\mathcal{F}_{l}(0)$, $a$, $b$, $c$, and $d$ characterize the momentum-transfer squared dependence of the $\Omega_{b}^{*}\rightarrow\Omega_{c}^{*}$ transition form factors associated with the different Lorentz structures. They are obtained by fitting the QCD sum rule results using the central values of the auxiliary parameters, as determined in our previous study, Ref.~\cite{Amiri:2025gcf}, where their numerical values are reported in Tables~3 and~4. The relevant form factors are subsequently evaluated at $q^2 = m_P^2$ and $q^2 = m_V^2$ for the processes with pseudoscalar and vector mesons in the final states, respectively. These values are then inserted into Eqs.~(\ref{DeWidPsem}) and~(\ref{DeWidVecm}) to determine the corresponding decay widths. The resulting numerical values for the form factors across the different decay channels are summarized in Tables~\ref{tab:PseforfacV},~\ref{tab:PseforfacAV},~\ref{tab:VecforfacV}, and~\ref{tab:VecforfacAV}.
%%%%%%%%%%%%%%%%%%%%%%%%%%%%%%%%%%%%%%%%%%%
\begin{table}[tbp]
	\centering
	\begin{tabular}{|c|c|c|c|c|}
		\hline\hline
		Process& $F_1(m_{P}^2)$ & $F_3(m_{P}^2)$ & $F_4(m_{P}^2)$ & $F_6(m_{P}^2)$ \\ \hline\hline
		$\Omega_{b}^{*}\rightarrow\Omega_{c}^{*}\pi^-$ & $9.21\pm1.05$ & $-0.60\pm 0.05$ &$-3.27\pm0.32$ &$-0.85\pm0.08$ \\
		$\Omega_{b}^{*}\rightarrow\Omega_{c}^{*}K^-$ & $9.30\pm 1.06$ & $-0.60\pm0.05$& $-3.31\pm0.32$ & $-0.86\pm0.09$ \\
		$\Omega_{b}^{*}\rightarrow\Omega_{c}^{*}D^-$ & $10.83\pm1.23$ & $-0.70\pm0.06$& $-3.93\pm0.38$& $-1.09\pm0.11$ \\
		$\Omega_{b}^{*}\rightarrow\Omega_{c}^{*}D_s^-$ & $11.03\pm 1.25$ & $-0.71\pm0.06$& $-4.02\pm0.39$& $-1.12\pm0.12$ \\
		\hline\hline
	\end{tabular}%
	\caption{The vector form factors contributing to the processes with a pesudoscalar meson in the final state. }
	\label{tab:PseforfacV}
\end{table}
%%%%%%%%%%%%%%%%%%%%%%%%%%%%%%%%%%%%%%%%%%%
%%%%%%%%%%%%%%%%%%%%%%%%%%%%%%%%%%%%%%%%%%%
\begin{table}[tbp]
	\centering
	\begin{tabular}{|c|c|c|c|c|}
		\hline\hline
		Process & $G_1(m_{P}^2)$ & $G_3(m_{P}^2)$ & $G_4(m_{P}^2)$ & $G_6(m_{P}^2)$ \\ \hline\hline
		$\Omega_{b}^{*}\rightarrow\Omega_{c}^{*}\pi^-$ & $3.07\pm0.29$ & $-2.34\pm0.27$& $3.41\pm0.33$& $-3.16\pm0.33$ \\
		$\Omega_{b}^{*}\rightarrow\Omega_{c}^{*}K^-$ & $3.09\pm0.30 $ & $-2.36\pm0.28$& $3.45\pm0.33$ & $-3.21\pm0.34$ \\
		$\Omega_{b}^{*}\rightarrow\Omega_{c}^{*}D^-$ & $3.34\pm0.32$ & $-2.76\pm0.32$ & $4.03\pm0.40$ & $-4.00\pm0.42$ \\
		$\Omega_{b}^{*}\rightarrow\Omega_{c}^{*}D_s^-$ & $3.37\pm 0.33$ &$-2.81\pm0.32$ & $4.11\pm0.40$& $-4.11\pm0.43$ \\
		\hline\hline
	\end{tabular}%
	\caption{The axial vector form factors contributing to the processes with a pseudoscalar meson in the final state.}
	\label{tab:PseforfacAV}
\end{table}
%%%%%%%%%%%%%%%%%%%%%%%%%%%%%%%%%%%%%%%%%%%
%%%%%%%%%%%%%%%%%%%%%%%%%%%%%%%%%%%%%%%%%%%
\begin{table}[tbp]
	\centering
	\begin{tabular}{|c|c|c|c|c|c|}
		\hline\hline
		Process & $F_1(m_{V}^2)$ & $F_2(m_{V}^2)$ & $F_4(m_{V}^2)$ & $F_5(m_{V}^2)$ & $F_7(m_{V}^2)$ \\ \hline\hline
		$\Omega_{b}^{*}\rightarrow\Omega_{c}^{*}\rho^-$ & $9.45\pm1.08$ &$-3.57\pm0.45$& $-3.37\pm0.33$ & $3.02\pm0.30$ & $2.52\pm0.21$\\
		$\Omega_{b}^{*}\rightarrow\Omega_{c}^{*}K^{*-}$ & $9.54\pm1.08$ & $-3.60\pm0.46$& $-3.40\pm0.33$ & $3.05\pm0.30$& $2.54\pm0.21$\\
		$\Omega_{b}^{*}\rightarrow\Omega_{c}^{*}D^{*-}$ & $11.12\pm1.27$ & $-4.24\pm0.54$& $-4.06\pm0.39$ & $3.78\pm0.37$ & $2.91\pm0.24$\\
		$\Omega_{b}^{*}\rightarrow\Omega_{c}^{*}D_s^{*-}$ & $11.36\pm1.29$ & $-4.34\pm0.55$& $-4.16\pm0.40$& $3.89\pm0.38$& $2.97\pm0.25$ \\
		\hline\hline
	\end{tabular}%
	\caption{The vector form factors contributing to the processes with a vector meson in the final state. }
	\label{tab:VecforfacV}
\end{table}
%%%%%%%%%%%%%%%%%%%%%%%%%%%%%%%%%%%%%%%%%%%
%%%%%%%%%%%%%%%%%%%%%%%%%%%%%%%%%%%%%%%%%%%
\begin{table}[tbp]
	\centering
	\begin{tabular}{|c|c|c|c|c|c|}
		\hline\hline
		Process & $G_1(m_{V}^2)$ & $G_2(m_{V}^2)$ & $G_4(m_{V}^2)$ & $G_5(m_{V}^2)$ & $G_7(m_{V}^2)$ \\ \hline\hline
		$\Omega_{b}^{*}\rightarrow\Omega_{c}^{*}\rho^-$ & $3.12\pm0.30$  & $-2.71\pm0.30$& $3.51\pm0.33$ & $0.48\pm0.05$& $-0.27\pm0.05$\\
		$\Omega_{b}^{*}\rightarrow\Omega_{c}^{*}K^{*-}$ & $3.13\pm0.30$ & $-2.73\pm0.30$& $3.54\pm0.34$ &$0.48\pm0.05$ & $-0.27\pm0.05$\\
		$\Omega_{b}^{*}\rightarrow\Omega_{c}^{*}D^{*-}$ & $3.39\pm0.32$ & $-3.24\pm0.36$& $4.14\pm0.40$ & $0.58\pm0.06$& $-0.30\pm0.06$ \\
		$\Omega_{b}^{*}\rightarrow\Omega_{c}^{*}D_s^{*-}$ &$3.42\pm0.33$ &$-3.31\pm0.37$ &$4.23\pm0.41$ &$0.59\pm0.07$ &$-0.31\pm0.06$\\
		\hline\hline
	\end{tabular}%
	\caption{The axial vector form factors contributing to the processes with a vector meson in the final state. }
	\label{tab:VecforfacAV}
\end{table}
%%%%%%%%%%%%%%%%%%%%%%%%%%%%%%%%%%%%%%%%%%%
\subsection{$\Omega_{b}^{*}\rightarrow\Omega_{c}^{*}P$ decays}\label{Pseudo}
In these decay channels, the final mesonic state is a pseudoscalar meson, specifically $\pi^-$, $K^-$, $D^-$ and $D_s^-$. The decay widths were computed using the input parameters and form factors listed in Tables~\ref{tab:MesonPar1},~\ref{tab:PseforfacV}, and~\ref{tab:PseforfacAV}, together with Eqs.~(\ref{eq:SquAmPseM}) and~(\ref{DeWidPsem}). The obtained results are presented in Table~\ref{tab:PseDecW}.
%%%%%%%%%%%%%%%%%%%%%%%%%%%%%%%%%%%%%%%%%%%
\begin{table}[tbp]
	\centering
	\begin{tabular}{|c|c|c|c|c|}
		\hline\hline
		Process& $\Omega_{b}^{*}\rightarrow\Omega_{c}^{*}\pi^-$ & $\Omega_{b}^{*}\rightarrow\Omega_{c}^{*}K^-$ & $\Omega_{b}^{*}\rightarrow\Omega_{c}^{*}D^-$ & $\Omega_{b}^{*}\rightarrow\Omega_{c}^{*}D_s^-$ \\ \hline\hline
		 $\Gamma_i$(GeV)& $5.59_{-1.04}^{+1.33}\times10^{-13}$ & $4.18_{-0.78}^{+0.99}\times10^{-14}$ &$6.39_{-1.23}^{+1.58}\times10^{-14}$ &$2.12_{-0.41}^{+0.53}\times10^{-12}$ \\
		\hline\hline
	\end{tabular}%
	\caption{The decay widths of the $\Omega_{b}^{*}\rightarrow\Omega_{c}^{*} P$ transitions. }
	\label{tab:PseDecW}
\end{table}
%%%%%%%%%%%%%%%%%%%%%%%%%%%%%%%%%%%%%%%%%%%
\subsection{$\Omega_{b}^{*}\rightarrow\Omega_{c}^{*}V$ decays}\label{Vector}
In these decay channels, the final mesonic state is a vector meson, namely $\rho^-$, $K^{*-}$, $D^{*-}$ and $D_s^{*-}$. The decay widths were evaluated using the input parameters and form factors listed in Tables~\ref{tab:MesonPar1},~\ref{tab:VecforfacV}, and~\ref{tab:VecforfacAV}, together with Eqs.~(\ref{eq:SquAmVecM}) and~(\ref{DeWidVecm}). The resulting decay widths are summarized in Table~\ref{tab:VecDecW}.
%%%%%%%%%%%%%%%%%%%%%%%%%%%%%%%%%%%%%%%%%%%
\begin{table}[tbp]
	\centering
	\begin{tabular}{|c|c|c|c|c|}
		\hline\hline
		Process& $\Omega_{b}^{*}\rightarrow\Omega_{c}^{*}\rho^-$ & $\Omega_{b}^{*}\rightarrow\Omega_{c}^{*}K^{*-}$ & $\Omega_{b}^{*}\rightarrow\Omega_{c}^{*}D^{*-}$ & $\Omega_{b}^{*}\rightarrow\Omega_{c}^{*}D_s^{*-}$ \\ \hline\hline
		$\Gamma_i$(GeV)& $1.47_{-0.28}^{+0.35}\times10^{-12}$ & $7.28_{-1.36}^{+1.75}\times10^{-14}$ &$6.44_{-1.21}^{+1.55}\times10^{-14}$ &$1.79_{-0.34}^{+0.42}\times10^{-12}$ \\
		\hline\hline
	\end{tabular}%
	\caption{The decay widths of the $\Omega_{b}^{*}\rightarrow\Omega_{c}^{*} V$ transitions. }
	\label{tab:VecDecW}
\end{table}
%%%%%%%%%%%%%%%%%%%%%%%%%%%%%%%%%%%%%%%%%%%
\subsection{$\Omega_{b}^{*}\rightarrow\Omega_{c}^{*}X$ decays}
As discussed in our previous work~~\cite{Amiri:2025gcf}, we performed a detailed analysis of the semileptonic weak decays $\Omega_{b}^{*}\rightarrow\Omega_{c}^{*}\ell\bar{\nu}_{\ell}$. In the present study, we adopt the decay width of these processes from that work and combine them with the results obtained in subsections~\ref{Pseudo} and~\ref{Vector} to analyze the dominant two-body nonleptonic decays of $\Omega_{b}^{*}\rightarrow\Omega_{c}^{*}$. The resulting decay widths for these channels are presented in Table~\ref{tab:AllDecW}. For the semileptonic channels, the total decay width is $6.79_{-1.21}^{+1.57}\times10^{-12}\ \text{GeV}$, while for the nonleptonic decay modes considered in this work it is $6.18_{-1.17}^{+1.49}\times10^{-12}\ \text{GeV}$. Among the nonleptonic transitions, the Cabibbo-suppressed processes $\Omega_{b}^{*}\rightarrow\Omega_{c}^{*}K^-$, $\Omega_{c}^{*}D^-$, $\Omega_{c}^{*}K^{*-}$ and $\Omega_{c}^{*}D^{*-}$ exhibit the smallest decay widths. Although other kinematically allowed two-body decays may exist,  they are expected to contribute subdominantly due to CKM suppression, flavor structure, and reduced phase space. In addition,  multibody decay channels are also possible but are further suppressed by smaller hadronic matrix elements and limited phase space. Therefore, the decay widths reported here capture the dominant contributions to the weak decays of $\Omega_b^*$. 
%%%%%%%%%%%%%%%%%%%%%%%%%%%%%%%%%%%%%%%%%%%
\begin{table}[tbp]
	\centering
	\begin{tabular}{|c|c|}
		\hline\hline
		Process&   $\Gamma_i$(GeV)   \\ \hline\hline
		$\Omega_{b}^{*}\rightarrow\Omega_{c}^{*} e \bar{\nu}_{e}$& $2.97_{-0.53}^{+0.69}\times10^{-12}$~\cite{Amiri:2025gcf} \\
		$\Omega_{b}^{*}\rightarrow\Omega_{c}^{*}\mu \bar{\nu}_{\mu}$& $2.96_{-0.53}^{+0.68}\times10^{-12}$~\cite{Amiri:2025gcf} \\
		$\Omega_{b}^{*}\rightarrow\Omega_{c}^{*}\tau \bar{\nu}_{\tau}$& $0.86_{-0.15}^{+0.20}\times10^{-12}$~\cite{Amiri:2025gcf} \\
		$\Omega_{b}^{*}\rightarrow\Omega_{c}^{*}\pi^-$ & $5.59_{-1.04}^{+1.33}\times10^{-13}$ \\
		$\Omega_{b}^{*}\rightarrow\Omega_{c}^{*}K^-$ & $4.18_{-0.78}^{+0.99}\times10^{-14}$ \\
		$\Omega_{b}^{*}\rightarrow\Omega_{c}^{*}D^-$ & $6.39_{-1.23}^{+1.58}\times10^{-14}$ \\
		$\Omega_{b}^{*}\rightarrow\Omega_{c}^{*}D_s^-$& $2.12_{-0.41}^{+0.53}\times10^{-12}$ \\
		$\Omega_{b}^{*}\rightarrow\Omega_{c}^{*}\rho^-$& $1.47_{-0.28}^{+0.35}\times10^{-12}$ \\
		$\Omega_{b}^{*}\rightarrow\Omega_{c}^{*}K^{*-}$ & $7.28_{-1.36}^{+1.75}\times10^{-14}$ \\
		$\Omega_{b}^{*}\rightarrow\Omega_{c}^{*}D^{*-}$ &$6.44_{-1.21}^{+1.55}\times10^{-14}$ \\
		$\Omega_{b}^{*}\rightarrow\Omega_{c}^{*}D_s^{*-}$& $1.79_{-0.34}^{+0.42}\times10^{-12}$ \\ \hline
		Total ($\Omega_{b}^{*}\rightarrow\Omega_{c}^{*}X$) & $12.97_{-2.38}^{+3.06}\times10^{-12}$ \\
		\hline\hline
	\end{tabular}%
	\caption{The decay widths of weak decay channels of the $\Omega_{b}^*$ baryon.}
	\label{tab:AllDecW}
\end{table}
%%%%%%%%%%%%%%%%%%%%%%%%%%%%%%%%%%%%%%%%%%%

%%%%%%%%%%%%%%%%%%%%%%%%%%%%%%%%%%%%%%%%%%%
\section{Conclusions}\label{sec:conclusions}
%%%%%%%%%%%%%%%%%%%%%%%%%%%%%%%%%%%%%%%%%%%%%%%
The experimental observation of the $\Omega_{b}^{*}$ baryon remains challenging, which underscores the importance of theoretical investigations of its weak decay channels. In this work, we have conducted a detailed analysis of the dominant two-body nonleptonic weak decays $\Omega_{b}^{*}\rightarrow\Omega_{c}^{*}P(V)$ within the factorization framework, a method that has proven reliable for studying two-body baryonic decays with a meson in the final state. Utilizing the $\Omega_{b}^{*}\rightarrow\Omega_{c}^{*}$ transition form factors obtained in Ref.~\cite{Amiri:2025gcf}, we have calculated the decay amplitudes and decay widths for the dominant nonleptonic channels. The numerical results are summarized in Table~\ref{tab:AllDecW}. 

For the complete set of weak decay modes considered here and in our earlier work~\cite{Amiri:2025gcf}, we obtained a total decay width of $12.97_{-2.38}^{+3.06}\times10^{-12}\ \text{GeV}$. Among these, the semileptonic decays account for $6.79_{-1.21}^{+1.57}\times10^{-12}\ \text{GeV}$, while nonleptonic channels contribute $6.18_{-1.17}^{+1.49}\times10^{-12}\ \text{GeV}$, highlighting the significant role of nonleptonic transitions in the overall weak decay dynamics of the $\Omega_{b}^{*}$ baryon. 

Although there may exist other two-body decays, they are expected to be subdominant due to CKM suppression, flavor structure, and limited phase space. Multibody decay channels are also possible but are further suppressed by smaller hadronic matrix elements and phase-space limitations. Therefore, the decay widths presented in this work capture the dominant contributions to the weak decays of $\Omega_{b}^{*}$.

Together with Ref.~\cite{Amiri:2025gcf}, the present study provides a theoretical framework and quantitative predictions that may assist and support the eventual experimental establishment of the $\Omega_{b}^{*}$ state.  
%%%%%%%%%%%%%%%%%%%%%%%%%%%%%%%%%%%%%%%%%%%%%
%\section*{Acknowledgements}
%%%%%%%%%%%%%%%%%%%%%%%%%%%%%%%%%%%%%%%%%%%%%
%%%%%%%%%%%%%%%%%%%%%%%%%%%%%%%%%%%%%%%%%%%%%
\appendix
%%%%%%%%%%%%%%%%%%%%%%%%%%%%%%%%%%%%%%%%%%%%%
%%%%%%%%%%%%%%%%%%%%%%%%%%%%%%%%%%%%%%%%%%%%%
\section{Amplitude squared} \label{Appen}
%%%%%%%%%%%%%%%%%%%%%%%%%%%%%%%%%%%%%%%%%%%%%
The squared amplitude for the processes with a pseudoscalar meson in the final state (i.e., the squared modulus of the amplitude $\mathcal{A}_P(\Omega_{b}^{*}\rightarrow\Omega_{c}^{*}P)$) is expressed as,
\begin{align} \label{eq:SquAmPseM}
    &|\mathcal{A}_P|^2 = \frac{1}{2} G_F^2 |V_{bc}|^2 |V_{qq'}|^2 a_1^2(\mu) f_P^2 \Big\{\nonumber \\&
    -\Big(-\frac{m_P^2}{m_{\Omega_b^*}} G_3(m_P^2)+ m_+\, G_1(m_P^2)\Big)^2 
    \Big[-\frac{28\, m_a}{9} - \frac{2 m_a^3}{9 m_b^2} + \frac{8 m_a^2}{9 m_b} + \frac{40 m_b}{9}\Big]\nonumber \\& \quad + \Big(\frac{m_P^2}{m_{\Omega_b^*}}F_3(m_P^2)  + m_-\, F_1(m_P^2)\Big)^2
    \Big[\frac{28\, m_a}{9} + \frac{2 m_a^3}{9 m_b^2} + \frac{8 m_a^2}{9 m_b} + \frac{40 m_b}{9}\Big]\nonumber \\& \quad - 2\Big(\frac{m_P^2}{m_{\Omega_b^*}}F_3(m_P^2) + m_- F_1(m_P^2) \Big)
    \Big(\frac{m_P^2}{m_{\Omega_b^*}^3}F_6(m_P^2) + \frac{m_-}{m_{\Omega_b^*}^2}F_4(m_P^2)\Big)
    \Big[-\frac{2 m_a^2}{9} + \frac{m_a^4}{9 m_b^2} + \frac{m_a^3}{3 m_b}\nonumber \\& - \frac{4 m_a m_b}{3} - \frac{8 m_b^2}{9}\Big] + 2\Big(-\frac{m_P^2}{m_{\Omega_b^*}}G_3(m_P^2) + m_+\, G_1(m_P^2)\Big)
    \Big(-\frac{m_P^2}{m_{\Omega_b^*}^3}G_6(m_P^2) + \frac{m_+}{m_{\Omega_b^*}^2}G_4(m_P^2)\Big)\nonumber \\&
    \times\Big[\frac{2 m_a^2}{9} - \frac{m_a^4}{9 m_b^2} + \frac{m_a^3}{3 m_b} - \frac{4 m_a m_b}{3} + \frac{8 m_b^2}{9}\Big]\nonumber \\& \quad - \Big(-\frac{m_P^2}{m_{\Omega_b^*}^3}G_6(m_P^2) + \frac{m_+}{m_{\Omega_b^*}^2}G_4(m_P^2)\Big)^2
    \Big[\frac{4 m_a^3}{9} - \frac{m_a^5}{18 m_b^2} + \frac{m_a^4}{9 m_b} - \frac{8 m_a^2 m_b}{9} - \frac{8 m_a m_b^2}{9} + \frac{16 m_b^3}{9}\Big]\nonumber \\& + \Big(\frac{m_P^2}{m_{\Omega_b^*}^3}F_6(m_P^2) + \frac{m_-}{m_{\Omega_b^*}^2}F_4(m_P^2)\Big)^2
    \Big[-\frac{4 m_a^3}{9} + \frac{m_a^5}{18 m_b^2} + \frac{m_a^4}{9 m_b} - \frac{8 m_a^2 m_b}{9} + \frac{8 m_a m_b^2}{9} + \frac{16 m_b^3}{9}\Big] \Big\} \,,
\end{align}
Similarly, the squared amplitude for the processes with a vector meson in the final state (i.e., the squared modulus of the amplitude $\mathcal{A}_V(\Omega_{b}^{*}\rightarrow\Omega_{c}^{*}V)$) is given by,
\begin{align} \label{eq:SquAmVecM}
		&|\mathcal{A}_V|^2 = \frac{1}{2} G_F^2 |V_{bc}|^2 |V_{qq'}|^2 a_1^2(\mu) m_V^2 f_V^2 \Big\{ \nonumber \\& 
		-\frac{2}{m_{\Omega_b^*}} F_7(m_V^2) \Big(F_1(m_V^2) + \frac{m_+}{m_{\Omega_b^*}}F_2(m_V^2)\Big)  
		\Big[
		\frac{32 m_{\Omega_b^*} m_b}{9} + \frac{32 m_{\Omega_c^*} m_b}{9} 
		+ \frac{16 m_{\Omega_b^*} m_c}{9} \nonumber \\& + \frac{16 m_{\Omega_c^*} m_c}{9} 
		- \frac{8 m_c^2}{9 m_{\Omega_b^*}} - \frac{8 m_c^2}{9 m_{\Omega_c^*}} - \frac{4 m_c^3}{9 m_{\Omega_b^*} m_b} - \frac{4 m_c^3}{9 m_{\Omega_c^*} m_b}
		\Big] \nonumber \\&
		+ \frac{2}{m_{\Omega_b^*}} G_7(m_V^2) \Big(-G_1(m_V^2) + \frac{m_-}{m_{\Omega_b^*}}G_2(m_V^2)\Big)
		\Big[
		-\frac{32 m_{\Omega_b^*} m_b}{9} + \frac{32 m_{\Omega_c^*} m_b}{9} 
		+ \frac{16 m_{\Omega_b^*} m_c}{9} \nonumber \\& - \frac{16 m_{\Omega_c^*} m_c}{9}- \frac{8 m_c^2}{9 m_{\Omega_b^*}} + \frac{8 m_c^2}{9 m_{\Omega_c^*}} 
		+ \frac{4 m_c^3}{9 m_{\Omega_b^*} m_b} - \frac{4 m_c^3}{9 m_{\Omega_c^*} m_b} \Big] \nonumber \\&
		+\Big(F_1(m_V^2) + \frac{m_+}{m_{\Omega_b^*}}F_2(m_V^2)\Big)^2 
		\Big[
		-\frac{128 m_b}{9} + \frac{40 m_{\Omega_b^*}^2 m_b}{9 m_V^2} + \frac{40 m_{\Omega_c^*}^2 m_b}{9 m_V^2} 
		- \frac{80 m_b^2}{9 m_V^2} + \frac{8 m_c}{9} \nonumber \\& + \frac{28 m_{\Omega_b^*}^2 m_c}{9 m_V^2} 
		+ \frac{28 m_{\Omega_c^*}^2 m_c}{9 m_V^2} - \frac{56 m_b m_c}{9 m_V^2} 
		- \frac{16 m_c^2}{9 m_V^2} + \frac{8 m_{\Omega_b^*} m_c^2}{9 m_{\Omega_c^*} m_V^2} + \frac{8 m_{\Omega_c^*} m_c^2}{9 m_{\Omega_b^*} m_V^2} 
		- \frac{16 m_c^2}{9 m_b} \nonumber \\& + \frac{2 m_c^3}{9 m_{\Omega_b^*}^2 m_V^2} + \frac{2 m_c^3}{9 m_{\Omega_c^*}^2 m_V^2} 
		+ \frac{4 m_c^3}{9 m_b^2} - \frac{4 m_c^3}{9 m_V^2 m_b}
		\Big] \nonumber \\&
		+ \Big(-G_1(m_V^2) + \frac{m_-}{m_{\Omega_b^*}}G_2(m_V^2)\Big)^2
		\Big[
		\frac{128 m_b}{9} - \frac{40 m_{\Omega_b^*}^2 m_b}{9 m_V^2} - \frac{40 m_{\Omega_c^*}^2 m_b}{9 m_V^2} 
		- \frac{80 m_b^2}{9 m_V^2} + \frac{8 m_c}{9} \nonumber \\& + \frac{28 m_{\Omega_b^*}^2 m_c}{9 m_V^2} 
		+ \frac{28 m_{\Omega_c^*}^2 m_c}{9 m_V^2} + \frac{56 m_b m_c}{9 m_V^2} 
		- \frac{16 m_c^2}{9 m_V^2}  - \frac{8 m_{\Omega_b^*} m_c^2}{9 m_{\Omega_c^*} m_V^2} - \frac{8 m_{\Omega_c^*} m_c^2}{9 m_{\Omega_b^*} m_V^2} 
		+ \frac{16 m_c^2}{9 m_b} \nonumber \\& + \frac{2 m_c^3}{9 m_{\Omega_b^*}^2 m_V^2} + \frac{2 m_c^3}{9 m_{\Omega_c^*}^2 m_V^2} 
		+ \frac{4 m_c^3}{9 m_b^2} + \frac{4 m_c^3}{9 m_V^2 m_b}
		\Big] \nonumber \\&
		+ \frac{1}{m_{\Omega_b^*}^2} F_7^2(m_V^2) 
		\Big[
		-\frac{16}{3} m_{\Omega_b^*}^2 m_b - \frac{16 m_{\Omega_c^*}^2 m_b}{3} - \frac{16 m_b^2}{9} 
		- \frac{8 m_{\Omega_b^*}^2 m_c}{3} - \frac{8 m_{\Omega_c^*}^2 m_c}{3} + \frac{8 m_b m_c}{9} \nonumber \\&
		+ \frac{4 m_c^2}{3} + \frac{4 m_{\Omega_b^*} m_c^2}{3 m_{\Omega_c^*}} + \frac{4 m_{\Omega_c^*} m_c^2}{3 m_{\Omega_b^*}} 
		+ \frac{2 m_c^3}{3 m_{\Omega_b^*}^2} + \frac{2 m_c^3}{3 m_{\Omega_c^*}^2} - \frac{2 m_c^3}{9 m_b} - \frac{2 m_c^4}{9 m_b^2}
		\Big] \nonumber \\&
		+ \frac{4}{m_{\Omega_b^*}^2}F_2(m_V^2) F_7(m_V^2) 
		\Big[
		\frac{8 m_b^2}{9} + \frac{4 m_b m_c}{3} + \frac{2 m_c^2}{9} - \frac{m_c^3}{3 m_b} - \frac{m_c^4}{9 m_b^2}
		\Big] \nonumber \\&
		- \frac{4}{m_{\Omega_b^*}^2} G_2(m_V^2) G_7(m_V^2)
		\Big[
		-\frac{8 m_b^2}{9} + \frac{4 m_b m_c}{3} - \frac{2 m_c^2}{9} - \frac{m_c^3}{3 m_b} + \frac{m_c^4}{9 m_b^2}
		\Big] \nonumber \\&
		- \frac{1}{m_{\Omega_b^*}^2} G_7^2(m_V^2)
		\Big[
		-\frac{16}{3} m_{\Omega_b^*}^2 m_b - \frac{16 m_{\Omega_c^*}^2 m_b}{3} + \frac{16 m_b^2}{9} 
		+ \frac{8 m_{\Omega_b^*}^2 m_c}{3} + \frac{8 m_{\Omega_c^*}^2 m_c}{3} + \frac{8 m_b m_c}{9} \nonumber \\&
		- \frac{4 m_c^2}{3} + \frac{4 m_{\Omega_b^*} m_c^2}{3 m_{\Omega_c^*}} + \frac{4 m_{\Omega_c^*} m_c^2}{3 m_{\Omega_b^*}} 
		- \frac{2 m_c^3}{3 m_{\Omega_b^*}^2} - \frac{2 m_c^3}{3 m_{\Omega_c^*}^2} - \frac{2 m_c^3}{9 m_b} + \frac{2 m_c^4}{9 m_b^2} \Big] \nonumber \\&
		+ \frac{2}{m_{\Omega_b^*}} F_7(m_V^2) 
		\Big(
		\frac{1}{m_{\Omega_b^*}^2}F_4(m_V^2) + \frac{m_+}{m_{\Omega_b^*}^3}F_5(m_V^2)
		\Big)
		\Big[
			-\frac{32}{9} m_{\Omega_b^*} m_b^2 - \frac{32 m_{\Omega_c^*} m_b^2}{9} 
			+ \frac{16 m_{\Omega_b^*} m_c^2}{9} \nonumber \\& + \frac{16 m_{\Omega_c^*} m_c^2}{9} 
			- \frac{2 m_c^4}{9 m_{\Omega_b^*} m_b} - \frac{2 m_c^4}{9 m_{\Omega_c^*} m_b} \Big] \nonumber \\&
		+ \frac{2}{m_{\Omega_b^*}} G_7(m_V^2)  
		\Big(
		\frac{1}{m_{\Omega_b^*}^2}G_4(m_V^2) - \frac{m_-}{m_{\Omega_b^*}^3}G_5(m_V^2)
		\Big)
		\Big[
			-\frac{32}{9} m_{\Omega_b^*} m_b^2 + \frac{32 m_{\Omega_c^*} m_b^2}{9} 
			+ \frac{16 m_{\Omega_b^*} m_c^2}{9} \nonumber \\& - \frac{16 m_{\Omega_c^*} m_c^2}{9} 
			+ \frac{2 m_c^4}{9 m_{\Omega_b^*} m_b} - \frac{2 m_c^4}{9 m_{\Omega_c^*} m_b} \Big] \nonumber \\& 
		- 2 \Big(\frac{1}{m_{\Omega_b^*}^2}F_4(m_V^2) + \frac{m_+}{m_{\Omega_b^*}^3}F_5(m_V^2)\Big)
		\Big(F_1(m_V^2) + \frac{m_+}{m_{\Omega_b^*}}F_2(m_V^2)\Big) 
		\Big[
		\frac{16 m_b^2}{9} - \frac{8 m_{\Omega_b^*}^2 m_b^2}{9 m_V^2} \nonumber \\& - \frac{8 m_{\Omega_c^*}^2 m_b^2}{9 m_V^2} 
		+ \frac{16 m_b^3}{9 m_V^2} + \frac{32 m_b m_c}{9} 
		- \frac{4 m_{\Omega_b^*}^2 m_b m_c}{3 m_V^2} - \frac{4 m_{\Omega_c^*}^2 m_b m_c}{3 m_V^2} + \frac{8 m_b^2 m_c}{3 m_V^2} 
		- \frac{4 m_c^2}{3} \nonumber \\& - \frac{2 m_{\Omega_b^*}^2 m_c^2}{9 m_V^2} - \frac{2 m_{\Omega_c^*}^2 m_c^2}{9 m_V^2} 
		+ \frac{4 m_b m_c^2}{9 m_V^2} 
		- \frac{2 m_c^3}{3 m_V^2} + \frac{m_{\Omega_b^*} m_c^3}{3 m_{\Omega_c^*} m_V^2} + \frac{m_{\Omega_c^*} m_c^3}{3 m_{\Omega_b^*} m_V^2} 
		- \frac{8 m_c^3}{9 m_b} \nonumber \\& + \frac{m_c^4}{9 m_{\Omega_b^*}^2 m_V^2} + \frac{m_c^4}{9 m_{\Omega_c^*}^2 m_V^2} 
		+ \frac{2 m_c^4}{9 m_b^2} - \frac{2 m_c^4}{9 m_V^2 m_b}
		\Big] \nonumber \\&
		+ 2 \Big(\frac{1}{m_{\Omega_b^*}^2}G_4(m_V^2) - \frac{m_-}{m_{\Omega_b^*}^3}G_5(m_V^2)\Big)
		\Big(-G_1(m_V^2) + \frac{m_-}{m_{\Omega_b^*}}G_2(m_V^2)\Big)
		\Big[
		\frac{16 m_b^2}{9} - \frac{8 m_{\Omega_b^*}^2 m_b^2}{9 m_V^2} \nonumber \\& - \frac{8 m_{\Omega_c^*}^2 m_b^2}{9 m_V^2} 
		- \frac{16 m_b^3}{9 m_V^2} - \frac{32 m_b m_c}{9} 
		+ \frac{4 m_{\Omega_b^*}^2 m_b m_c}{3 m_V^2} + \frac{4 m_{\Omega_c^*}^2 m_b m_c}{3 m_V^2} + \frac{8 m_b^2 m_c}{3 m_V^2} 
		- \frac{4 m_c^2}{3} \nonumber \\& - \frac{2 m_{\Omega_b^*}^2 m_c^2}{9 m_V^2} - \frac{2 m_{\Omega_c^*}^2 m_c^2}{9 m_V^2} 
		- \frac{4 m_b m_c^2}{9 m_V^2} - \frac{2 m_c^3}{3 m_V^2} 
		- \frac{m_{\Omega_b^*} m_c^3}{3 m_{\Omega_c^*} m_V^2} - \frac{m_{\Omega_c^*} m_c^3}{3 m_{\Omega_b^*} m_V^2} 
		+ \frac{8 m_c^3}{9 m_b} \nonumber \\& + \frac{m_c^4}{9 m_{\Omega_b^*}^2 m_V^2} + \frac{m_c^4}{9 m_{\Omega_c^*}^2 m_V^2} 
		+ \frac{2 m_c^4}{9 m_b^2} + \frac{2 m_c^4}{9 m_V^2 m_b}
		\Big]  \nonumber \\& 
		+\frac{4}{m_{\Omega_b^*}} G_2(m_V^2)
		\Big(-G_1(m_V^2) + \frac{m_-}{m_{\Omega_b^*}}G_2(m_V^2)\Big)
		\Big[
		\frac{40 m_{\Omega_c^*} m_b}{9} - \frac{40 m_{\Omega_c^*}^3 m_b}{9 m_V^2} - \frac{40 m_{\Omega_c^*} m_b^2}{9 m_V^2} \nonumber \\& 
		- \frac{28 m_{\Omega_c^*} m_c}{9} + \frac{28 m_{\Omega_c^*}^3 m_c}{9 m_V^2} + \frac{20 m_{\Omega_b^*} m_b m_c}{9 m_V^2} 
		+ \frac{16 m_{\Omega_c^*} m_b m_c}{3 m_V^2} + \frac{8 m_c^2}{9 m_{\Omega_b^*}} - \frac{14 m_{\Omega_b^*} m_c^2}{9 m_V^2} 
		- \frac{22 m_{\Omega_c^*} m_c^2}{9 m_V^2} \nonumber \\& - \frac{8 m_{\Omega_c^*}^2 m_c^2}{9 m_{\Omega_b^*} m_V^2} + \frac{2 m_c^3}{3 m_{\Omega_b^*} m_V^2} 
		+ \frac{4 m_c^3}{9 m_{\Omega_c^*} m_V^2} + \frac{2 m_{\Omega_c^*} m_c^3}{9 m_{\Omega_b^*}^2 m_V^2} - \frac{2 m_c^3}{9 m_{\Omega_b^*} m_b} 
		- \frac{m_c^4}{9 m_{\Omega_b^*} m_V^2 m_b} - \frac{m_c^4}{9 m_{\Omega_c^*} m_V^2 m_b}
		\Big] \nonumber \\&
		- \frac{4}{m_{\Omega_b^*}}F_2(m_V^2) 
		\Big(F_1(m_V^2) + \frac{m_+}{m_{\Omega_b^*}}F_2(m_V^2)\Big)
		\Big[
		-\frac{40 m_{\Omega_c^*} m_b}{9} + \frac{40 m_{\Omega_c^*}^3 m_b}{9 m_V^2} - \frac{40 m_{\Omega_c^*} m_b^2}{9 m_V^2} \nonumber \\&
		- \frac{28 m_{\Omega_c^*} m_c}{9} + \frac{28 m_{\Omega_c^*}^3 m_c}{9 m_V^2} + \frac{20 m_{\Omega_b^*} m_b m_c}{9 m_V^2} 
		- \frac{16 m_{\Omega_c^*} m_b m_c}{3 m_V^2} - \frac{8 m_c^2}{9 m_{\Omega_b^*}} + \frac{14 m_{\Omega_b^*} m_c^2}{9 m_V^2} 
		- \frac{22 m_{\Omega_c^*} m_c^2}{9 m_V^2} \nonumber \\& + \frac{8 m_{\Omega_c^*}^2 m_c^2}{9 m_{\Omega_b^*} m_V^2} - \frac{2 m_c^3}{3 m_{\Omega_b^*} m_V^2}
		+ \frac{4 m_c^3}{9 m_{\Omega_c^*} m_V^2} + \frac{2 m_{\Omega_c^*} m_c^3}{9 m_{\Omega_b^*}^2 m_V^2} - \frac{2 m_c^3}{9 m_{\Omega_b^*} m_b} 
		- \frac{m_c^4}{9 m_{\Omega_b^*} m_V^2 m_b} + \frac{m_c^4}{9 m_{\Omega_c^*} m_V^2 m_b}
		\Big] \nonumber \\&   
		-  \frac{4}{m_{\Omega_b^*}^4} F_5(m_V^2) F_7(m_V^2)
		\Big[
		-\frac{16 m_b^3}{9} - \frac{8 m_b^2 m_c}{9} + \frac{8 m_b m_c^2}{9} 
		+ \frac{4 m_c^3}{9} - \frac{m_c^4}{9 m_b} - \frac{m_c^5}{18 m_b^2}
		\Big] \nonumber \\&
		+ \frac{4}{m_{\Omega_b^*}^4} G_7(m_V^2) G_5(m_V^2)
		\Big[
		-\frac{16 m_b^3}{9} + \frac{8 m_b^2 m_c}{9} + \frac{8 m_b m_c^2}{9} 
		- \frac{4 m_c^3}{9} - \frac{m_c^4}{9 m_b} + \frac{m_c^5}{18 m_b^2}
		\Big] \nonumber \\&  
		-\frac{4}{m_{\Omega_b^*}^2} G_2^2(m_V^2)
		\Big[
		-\frac{40 m_{\Omega_c^*}^2 m_b}{9} + \frac{40 m_{\Omega_c^*}^4 m_b}{9 m_V^2} 
		+ \frac{28 m_{\Omega_c^*}^2 m_c}{9} - \frac{28 m_{\Omega_c^*}^4 m_c}{9 m_V^2} 
		- \frac{40 m_{\Omega_c^*}^2 m_b m_c}{9 m_V^2} \nonumber \\&
		- \frac{8 m_{\Omega_c^*} m_c^2}{9 m_{\Omega_b^*}} 
		+ \frac{28 m_{\Omega_c^*}^2 m_c^2}{9 m_V^2} 
		+ \frac{8 m_{\Omega_c^*}^3 m_c^2}{9 m_{\Omega_b^*} m_V^2} 
		+ \frac{10 m_b m_c^2}{9 m_V^2} 
		+ \frac{2 m_c^3}{9 m_{\Omega_b^*}^2} - \frac{7 m_c^3}{9 m_V^2} 
		- \frac{8 m_{\Omega_c^*} m_c^3}{9 m_{\Omega_b^*} m_V^2} 
		- \frac{2 m_{\Omega_c^*}^2 m_c^3}{9 m_{\Omega_b^*}^2 m_V^2} \nonumber \\& 
		+ \frac{2 m_c^4}{9 m_{\Omega_b^*}^2 m_V^2} 
		+ \frac{2 m_c^4}{9 m_V^2 m_b} 
		- \frac{m_c^5}{18 m_V^2 m_b^2} 
		\Big] \nonumber \\&
		+ \frac{4}{m_{\Omega_b^*}^2} F_2^2(m_V^2)
		\Big[
		-\frac{40 m_{\Omega_c^*}^2 m_b}{9} + \frac{40 m_{\Omega_c^*}^4 m_b}{9 m_V^2} 
		- \frac{28 m_{\Omega_c^*}^2 m_c}{9} + \frac{28 m_{\Omega_c^*}^4 m_c}{9 m_V^2} 
		- \frac{40 m_{\Omega_c^*}^2 m_b m_c}{9 m_V^2} \nonumber \\&
		- \frac{8 m_{\Omega_c^*} m_c^2}{9 m_{\Omega_b^*}} 
		- \frac{28 m_{\Omega_c^*}^2 m_c^2}{9 m_V^2} 
		+ \frac{8 m_{\Omega_c^*}^3 m_c^2}{9 m_{\Omega_b^*} m_V^2} 
		+ \frac{10 m_b m_c^2}{9 m_V^2} 
		- \frac{2 m_c^3}{9 m_{\Omega_b^*}^2} + \frac{7 m_c^3}{9 m_V^2} 
		- \frac{8 m_{\Omega_c^*} m_c^3}{9 m_{\Omega_b^*} m_V^2} 
		+ \frac{2 m_{\Omega_c^*}^2 m_c^3}{9 m_{\Omega_b^*}^2 m_V^2} \nonumber \\& 
		- \frac{2 m_c^4}{9 m_{\Omega_b^*}^2 m_V^2} 
		+ \frac{2 m_c^4}{9 m_V^2 m_b} 
		+ \frac{m_c^5}{18 m_V^2 m_b^2} 
		\Big] \nonumber \\& 
		+ \Big(\frac{1}{m_{\Omega_b^*}^2}F_4(m_V^2) + \frac{m_+}{m_{\Omega_b^*}^3}F_5(m_V^2)\Big)^2 
		\Big[
		-\frac{64 m_b^3}{9} + \frac{16 m_{\Omega_b^*}^2 m_b^3}{9 m_V^2} 
		+ \frac{16 m_{\Omega_c^*}^2 m_b^3}{9 m_V^2} - \frac{32 m_b^4}{9 m_V^2} 
		+ \frac{16}{9} m_b^2 m_c \nonumber \\& + \frac{8 m_{\Omega_b^*}^2 m_b^2 m_c}{9 m_V^2} 
		+ \frac{8 m_{\Omega_c^*}^2 m_b^2 m_c}{9 m_V^2} 
		- \frac{16 m_b^3 m_c}{9 m_V^2} + \frac{32}{9} m_b m_c^2 
		- \frac{8 m_{\Omega_b^*}^2 m_b m_c^2}{9 m_V^2} - \frac{8 m_{\Omega_c^*}^2 m_b m_c^2}{9 m_V^2} \nonumber \\& 
		+ \frac{16 m_b^2 m_c^2}{9 m_V^2} - \frac{8 m_c^3}{9} - \frac{4 m_{\Omega_b^*}^2 m_c^3}{9 m_V^2} 
		- \frac{4 m_{\Omega_c^*}^2 m_c^3}{9 m_V^2} + \frac{8 m_b m_c^3}{9 m_V^2} 
		- \frac{2 m_c^4}{9 m_V^2} + \frac{m_{\Omega_b^*} m_c^4}{9 m_{\Omega_c^*} m_V^2} 
		+ \frac{m_{\Omega_c^*} m_c^4}{9 m_{\Omega_b^*} m_V^2} \nonumber \\& - \frac{4 m_c^4}{9 m_b} 
		+ \frac{m_c^5}{18 m_{\Omega_b^*}^2 m_V^2} + \frac{m_c^5}{18 m_{\Omega_c^*}^2 m_V^2} 
		+ \frac{m_c^5}{9 m_b^2} - \frac{m_c^5}{9 m_V^2 m_b}
		\Big] \nonumber \\&
		+ \Big(\frac{1}{m_{\Omega_b^*}^2}G_4(m_V^2) - \frac{m_-}{m_{\Omega_b^*}^3}G_5(m_V^2)\Big)^2 
		\Big[
		\frac{64 m_b^3}{9} - \frac{16 m_{\Omega_b^*}^2 m_b^3}{9 m_V^2} 
		- \frac{16 m_{\Omega_c^*}^2 m_b^3}{9 m_V^2} - \frac{32 m_b^4}{9 m_V^2} 
		+ \frac{16}{9} m_b^2 m_c \nonumber \\& + \frac{8 m_{\Omega_b^*}^2 m_b^2 m_c}{9 m_V^2} 
		+ \frac{8 m_{\Omega_c^*}^2 m_b^2 m_c}{9 m_V^2} 
		+ \frac{16 m_b^3 m_c}{9 m_V^2} - \frac{32}{9} m_b m_c^2 
		+ \frac{8 m_{\Omega_b^*}^2 m_b m_c^2}{9 m_V^2} + \frac{8 m_{\Omega_c^*}^2 m_b m_c^2}{9 m_V^2} \nonumber \\&
		+ \frac{16 m_b^2 m_c^2}{9 m_V^2} - \frac{8 m_c^3}{9} - \frac{4 m_{\Omega_b^*}^2 m_c^3}{9 m_V^2} 
		- \frac{4 m_{\Omega_c^*}^2 m_c^3}{9 m_V^2} - \frac{8 m_b m_c^3}{9 m_V^2}
		- \frac{2 m_c^4}{9 m_V^2} - \frac{m_{\Omega_b^*} m_c^4}{9 m_{\Omega_c^*} m_V^2} 
		- \frac{m_{\Omega_c^*} m_c^4}{9 m_{\Omega_b^*} m_V^2} \nonumber \\& + \frac{4 m_c^4}{9 m_b} 
		+ \frac{m_c^5}{18 m_{\Omega_b^*}^2 m_V^2} + \frac{m_c^5}{18 m_{\Omega_c^*}^2 m_V^2} 
		+ \frac{m_c^5}{9 m_b^2} + \frac{m_c^5}{9 m_V^2 m_b}
		\Big] \nonumber \\&
		+ \frac{4}{m_{\Omega_b^*}} G_2(m_V^2) 
		\Big(\frac{1}{m_{\Omega_b^*}^2}G_4(m_V^2) - \frac{m_-}{m_{\Omega_b^*}^3}G_5(m_V^2)\Big) 
		\Big[
		\frac{8 m_{\Omega_c^*} m_b^2}{9} - \frac{8 m_{\Omega_c^*}^3 m_b^2}{9 m_V^2} 
		- \frac{8 m_{\Omega_c^*} m_b^3}{9 m_V^2} \nonumber \\& - \frac{4}{3} m_{\Omega_c^*} m_b m_c 
		+ \frac{4 m_{\Omega_c^*}^3 m_b m_c}{3 m_V^2}
		+ \frac{4 m_{\Omega_b^*} m_b^2 m_c}{9 m_V^2} + \frac{16 m_{\Omega_c^*} m_b^2 m_c}{9 m_V^2} 
		+ \frac{2 m_{\Omega_c^*} m_c^2}{9} - \frac{2 m_{\Omega_c^*}^3 m_c^2}{9 m_V^2}  \nonumber \\& 
		- \frac{2 m_{\Omega_b^*} m_b m_c^2}{3 m_V^2} - \frac{8 m_{\Omega_c^*} m_b m_c^2}{9 m_V^2} 
		+ \frac{m_c^3}{3 m_{\Omega_b^*}} + \frac{m_{\Omega_b^*} m_c^3}{9 m_V^2} 
		- \frac{2 m_{\Omega_c^*} m_c^3}{9 m_V^2} - \frac{m_{\Omega_c^*}^2 m_c^3}{3 m_{\Omega_b^*} m_V^2} \nonumber \\&
		+ \frac{5 m_c^4}{18 m_{\Omega_b^*} m_V^2} + \frac{m_c^4}{6 m_{\Omega_c^*} m_V^2} 
		+ \frac{m_{\Omega_c^*} m_c^4}{9 m_{\Omega_b^*}^2 m_V^2} - \frac{m_c^4}{9 m_{\Omega_b^*} m_b} 
		- \frac{m_c^5}{18 m_{\Omega_b^*} m_V^2 m_b} - \frac{m_c^5}{18 m_{\Omega_c^*} m_V^2 m_b}
		\Big] \nonumber \\& 
		+ \frac{4}{m_{\Omega_b^*}^{3}} G_5(m_V^2) 
		\Big(-G_1(m_V^2) + \frac{m_-}{m_{\Omega_b^*}}G_2(m_V^2)\Big) 
		\Big[
		\frac{8 m_{\Omega_c^*} m_b^2}{9} - \frac{8 m_{\Omega_c^*}^3 m_b^2}{9 m_V^2} 
		- \frac{8 m_{\Omega_c^*} m_b^3}{9 m_V^2} \nonumber \\&  - \frac{4}{3} m_{\Omega_c^*} m_b m_c 
		+ \frac{4 m_{\Omega_c^*}^3 m_b m_c}{3 m_V^2} 
		+ \frac{4 m_{\Omega_b^*} m_b^2 m_c}{9 m_V^2} + \frac{16 m_{\Omega_c^*} m_b^2 m_c}{9 m_V^2} 
		+ \frac{2 m_{\Omega_c^*} m_c^2}{9} - \frac{2 m_{\Omega_c^*}^3 m_c^2}{9 m_V^2} \nonumber \\&  
		- \frac{2 m_{\Omega_b^*} m_b m_c^2}{3 m_V^2} - \frac{8 m_{\Omega_c^*} m_b m_c^2}{9 m_V^2} 
		+ \frac{m_c^3}{3 m_{\Omega_b^*}} + \frac{m_{\Omega_b^*} m_c^3}{9 m_V^2} 
		- \frac{2 m_{\Omega_c^*} m_c^3}{9 m_V^2} - \frac{m_{\Omega_c^*}^2 m_c^3}{3 m_{\Omega_b^*} m_V^2} \nonumber \\& 
		+ \frac{5 m_c^4}{18 m_{\Omega_b^*} m_V^2} + \frac{m_c^4}{6 m_{\Omega_c^*} m_V^2}
		+ \frac{m_{\Omega_c^*} m_c^4}{9 m_{\Omega_b^*}^2 m_V^2} - \frac{m_c^4}{9 m_{\Omega_b^*} m_b} 
		- \frac{m_c^5}{18 m_{\Omega_b^*} m_V^2 m_b} - \frac{m_c^5}{18 m_{\Omega_c^*} m_V^2 m_b}
		\Big] \nonumber \\& 
		+ \frac{4}{m_{\Omega_b^*}} F_2(m_V^2) \Big(\frac{1}{m_{\Omega_b^*}^2}F_4(m_V^2) + \frac{m_+}{m_{\Omega_b^*}^3}F_5(m_V^2)\Big) 
		\Big[
		\frac{8 m_{\Omega_c^*} m_b^2}{9} - \frac{8 m_{\Omega_c^*}^3 m_b^2}{9 m_V^2} 
		+ \frac{8 m_{\Omega_c^*} m_b^3}{9 m_V^2}  \nonumber \\&  + \frac{4}{3} m_{\Omega_c^*} m_b m_c 
		- \frac{4 m_{\Omega_c^*}^3 m_b m_c}{3 m_V^2}
		- \frac{4 m_{\Omega_b^*} m_b^2 m_c}{9 m_V^2} + \frac{16 m_{\Omega_c^*} m_b^2 m_c}{9 m_V^2} 
		+ \frac{2 m_{\Omega_c^*} m_c^2}{9} - \frac{2 m_{\Omega_c^*}^3 m_c^2}{9 m_V^2}  \nonumber \\& 
		- \frac{2 m_{\Omega_b^*} m_b m_c^2}{3 m_V^2} + \frac{8 m_{\Omega_c^*} m_b m_c^2}{9 m_V^2} 
		- \frac{m_c^3}{3 m_{\Omega_b^*}} - \frac{m_{\Omega_b^*} m_c^3}{9 m_V^2} - \frac{2 m_{\Omega_c^*} m_c^3}{9 m_V^2} 
		+ \frac{m_{\Omega_c^*}^2 m_c^3}{3 m_{\Omega_b^*} m_V^2} \nonumber \\& - \frac{5 m_c^4}{18 m_{\Omega_b^*} m_V^2} 
		+ \frac{m_c^4}{6 m_{\Omega_c^*} m_V^2} + \frac{m_{\Omega_c^*} m_c^4}{9 m_{\Omega_b^*}^2 m_V^2} 
		- \frac{m_c^4}{9 m_{\Omega_b^*} m_b} - \frac{m_c^5}{18 m_{\Omega_b^*} m_V^2 m_b} + \frac{m_c^5}{18 m_{\Omega_c^*} m_V^2 m_b}
		\Big] \nonumber \\& + \frac{4}{m_{\Omega_b^*}^{3}} F_5(m_V^2) \Big(F_1(m_V^2) + \frac{m_+}{m_{\Omega_b^*}}F_2(m_V^2)\Big) 
		\Big[
		\frac{8 m_{\Omega_c^*} m_b^2}{9} - \frac{8 m_{\Omega_c^*}^3 m_b^2}{9 m_V^2} 
		+ \frac{8 m_{\Omega_c^*} m_b^3}{9 m_V^2} \nonumber \\&  + \frac{4}{3} m_{\Omega_c^*} m_b m_c 
		- \frac{4 m_{\Omega_c^*}^3 m_b m_c}{3 m_V^2} 
		- \frac{4 m_{\Omega_b^*} m_b^2 m_c}{9 m_V^2} + \frac{16 m_{\Omega_c^*} m_b^2 m_c}{9 m_V^2} 
		+ \frac{2 m_{\Omega_c^*} m_c^2}{9} - \frac{2 m_{\Omega_c^*}^3 m_c^2}{9 m_V^2} \nonumber \\& 
		- \frac{2 m_{\Omega_b^*} m_b m_c^2}{3 m_V^2} + \frac{8 m_{\Omega_c^*} m_b m_c^2}{9 m_V^2} 
		- \frac{m_c^3}{3 m_{\Omega_b^*}} - \frac{m_{\Omega_b^*} m_c^3}{9 m_V^2} - \frac{2 m_{\Omega_c^*} m_c^3}{9 m_V^2} 
		+ \frac{m_{\Omega_c^*}^2 m_c^3}{3 m_{\Omega_b^*} m_V^2} \nonumber \\& - \frac{5 m_c^4}{18 m_{\Omega_b^*} m_V^2} 
		+ \frac{m_c^4}{6 m_{\Omega_c^*} m_V^2} + \frac{m_{\Omega_c^*} m_c^4}{9 m_{\Omega_b^*}^2 m_V^2} 
		- \frac{m_c^4}{9 m_{\Omega_b^*} m_b} - \frac{m_c^5}{18 m_{\Omega_b^*} m_V^2 m_b} + \frac{m_c^5}{18 m_{\Omega_c^*} m_V^2 m_b}
		\Big] \nonumber \\&
		+ \frac{8}{m_{\Omega_b^*}^4} G_2(m_V^2) G_5(m_V^2)
		\Big[
		- \frac{8}{9} m_{\Omega_c^*}^2 m_b^2 + \frac{8 m_{\Omega_c^*}^4 m_b^2}{9 m_V^2} 
		+ \frac{4}{3} m_{\Omega_c^*}^2 m_b m_c - \frac{4 m_{\Omega_c^*}^4 m_b m_c}{3 m_V^2} 
		- \frac{8 m_{\Omega_c^*}^2 m_b^2 m_c}{9 m_V^2} \nonumber \\& 
		- \frac{2}{9} m_{\Omega_c^*}^2 m_c^2 + \frac{2 m_{\Omega_c^*}^4 m_c^2}{9 m_V^2} 
		+ \frac{4 m_{\Omega_c^*}^2 m_b m_c^2}{3 m_V^2} + \frac{2 m_b^2 m_c^2}{9 m_V^2} 
		- \frac{m_{\Omega_c^*} m_c^3}{3 m_{\Omega_b^*}} - \frac{2 m_{\Omega_c^*}^2 m_c^3}{9 m_V^2}  
		+ \frac{m_{\Omega_c^*}^3 m_c^3}{3 m_{\Omega_b^*} m_V^2} - \frac{m_b m_c^3}{3 m_V^2} \nonumber \\& 
		+ \frac{m_c^4}{9 m_{\Omega_b^*}^2} + \frac{m_c^4}{18 m_V^2} 
		- \frac{m_{\Omega_c^*} m_c^4}{3 m_{\Omega_b^*} m_V^2} - \frac{m_{\Omega_c^*}^2 m_c^4}{9 m_{\Omega_b^*}^2 m_V^2} 
		+ \frac{m_c^5}{9 m_{\Omega_b^*}^2 m_V^2} + \frac{m_c^5}{12 m_V^2 m_b} - \frac{m_c^6}{36 m_V^2 m_b^2}
		\Big] \nonumber \\& - \frac{8}{m_{\Omega_b^*}^{4}} F_2(m_V^2) F_5(m_V^2)
		\Big[
		\frac{8}{9} m_{\Omega_c^*}^2 m_b^2 - \frac{8 m_{\Omega_c^*}^4 m_b^2}{9 m_V^2} 
		+ \frac{4}{3} m_{\Omega_c^*}^2 m_b m_c - \frac{4 m_{\Omega_c^*}^4 m_b m_c}{3 m_V^2} 
		+ \frac{8 m_{\Omega_c^*}^2 m_b^2 m_c}{9 m_V^2} \nonumber \\&
		+ \frac{2}{9} m_{\Omega_c^*}^2 m_c^2 - \frac{2 m_{\Omega_c^*}^4 m_c^2}{9 m_V^2} 
		+ \frac{4 m_{\Omega_c^*}^2 m_b m_c^2}{3 m_V^2} - \frac{2 m_b^2 m_c^2}{9 m_V^2} 
		- \frac{m_{\Omega_c^*} m_c^3}{3 m_{\Omega_b^*}} + \frac{2 m_{\Omega_c^*}^2 m_c^3}{9 m_V^2}
		+ \frac{m_{\Omega_c^*}^3 m_c^3}{3 m_{\Omega_b^*} m_V^2} - \frac{m_b m_c^3}{3 m_V^2}  \nonumber \\&
		- \frac{m_c^4}{9 m_{\Omega_b^*}^2} - \frac{m_c^4}{18 m_V^2} - \frac{m_{\Omega_c^*} m_c^4}{3 m_{\Omega_b^*} m_V^2} 
		+ \frac{m_{\Omega_c^*}^2 m_c^4}{9 m_{\Omega_b^*}^2 m_V^2} 
		- \frac{m_c^5}{9 m_{\Omega_b^*}^2 m_V^2} + \frac{m_c^5}{12 m_V^2 m_b} 
		+ \frac{m_c^6}{36 m_V^2 m_b^2} \Big] \nonumber \\& 
		+ \frac{4}{m_{\Omega_b^*}^3} G_5(m_V^2)
		\Big( \frac{1}{m_{\Omega_b^*}^2}G_4(m_V^2) - \frac{m_-}{m_{\Omega_b^*}^3}G_5(m_V^2) \Big)
		\Big[
		\frac{16}{9} m_{\Omega_c^*} m_b^3 - \frac{16 m_{\Omega_c^*}^3 m_b^3}{9 m_V^2} 
		- \frac{16 m_{\Omega_c^*} m_b^4}{9 m_V^2} \nonumber \\& - \frac{8}{9} m_{\Omega_c^*} m_b^2 m_c  
		+ \frac{8 m_{\Omega_c^*}^3 m_b^2 m_c}{9 m_V^2} + \frac{8 m_{\Omega_b^*} m_b^3 m_c}{9 m_V^2} 
		+ \frac{16 m_{\Omega_c^*} m_b^3 m_c}{9 m_V^2} - \frac{8}{9} m_{\Omega_c^*} m_b m_c^2 
		+ \frac{8 m_{\Omega_c^*}^3 m_b m_c^2}{9 m_V^2} \nonumber \\&
		- \frac{4 m_{\Omega_b^*} m_b^2 m_c^2}{9 m_V^2} + \frac{4 m_{\Omega_c^*} m_b^2 m_c^2}{9 m_V^2} 
		+ \frac{4 m_{\Omega_c^*} m_c^3}{9} - \frac{4 m_{\Omega_c^*}^3 m_c^3}{9 m_V^2} 
		- \frac{4 m_{\Omega_b^*} m_b m_c^3}{9 m_V^2} - \frac{8 m_{\Omega_c^*} m_b m_c^3}{9 m_V^2} \nonumber \\&
		+ \frac{m_c^4}{9 m_{\Omega_b^*}} + \frac{2 m_{\Omega_b^*} m_c^4}{9 m_V^2} + \frac{m_{\Omega_c^*} m_c^4}{9 m_V^2} 
		- \frac{m_{\Omega_c^*}^2 m_c^4}{9 m_{\Omega_b^*} m_V^2} + \frac{m_c^5}{9 m_{\Omega_b^*} m_V^2} 
		+ \frac{m_c^5}{18 m_{\Omega_c^*} m_V^2} + \frac{m_{\Omega_c^*} m_c^5}{18 m_{\Omega_b^*}^2 m_V^2} \nonumber \\&
		- \frac{m_c^5}{18 m_{\Omega_b^*} m_b} - \frac{m_c^6}{36 m_{\Omega_b^*} m_V^2 m_b} 
		- \frac{m_c^6}{36 m_{\Omega_c^*} m_V^2 m_b}
		\Big] \nonumber \\& - \frac{4}{m_{\Omega_b^*}^{3}} F_5(m_V^2)
		\Big( \frac{1}{m_{\Omega_b^*}^2}F_4(m_V^2) + \frac{m_+}{m_{\Omega_b^*}^3}F_5(m_V^2) \Big)
		\Big[
		-\frac{16}{9} m_{\Omega_c^*} m_b^3 + \frac{16 m_{\Omega_c^*}^3 m_b^3}{9 m_V^2} 
		- \frac{16 m_{\Omega_c^*} m_b^4}{9 m_V^2} \nonumber \\& - \frac{8}{9} m_{\Omega_c^*} m_b^2 m_c  
		+ \frac{8 m_{\Omega_c^*}^3 m_b^2 m_c}{9 m_V^2} + \frac{8 m_{\Omega_b^*} m_b^3 m_c}{9 m_V^2} 
		- \frac{16 m_{\Omega_c^*} m_b^3 m_c}{9 m_V^2} + \frac{8}{9} m_{\Omega_c^*} m_b m_c^2 
		- \frac{8 m_{\Omega_c^*}^3 m_b m_c^2}{9 m_V^2} \nonumber \\&
		+ \frac{4 m_{\Omega_b^*} m_b^2 m_c^2}{9 m_V^2} + \frac{4 m_{\Omega_c^*} m_b^2 m_c^2}{9 m_V^2} 
		+ \frac{4 m_{\Omega_c^*} m_c^3}{9} - \frac{4 m_{\Omega_c^*}^3 m_c^3}{9 m_V^2} 
		- \frac{4 m_{\Omega_b^*} m_b m_c^3}{9 m_V^2} + \frac{8 m_{\Omega_c^*} m_b m_c^3}{9 m_V^2} \nonumber \\&
		- \frac{m_c^4}{9 m_{\Omega_b^*}} - \frac{2 m_{\Omega_b^*} m_c^4}{9 m_V^2} + \frac{m_{\Omega_c^*} m_c^4}{9 m_V^2} 
		+ \frac{m_{\Omega_c^*}^2 m_c^4}{9 m_{\Omega_b^*} m_V^2} - \frac{m_c^5}{9 m_{\Omega_b^*} m_V^2} 
		+ \frac{m_c^5}{18 m_{\Omega_c^*} m_V^2} + \frac{m_{\Omega_c^*} m_c^5}{18 m_{\Omega_b^*}^2 m_V^2} \nonumber \\&
		- \frac{m_c^5}{18 m_{\Omega_b^*} m_b} - \frac{m_c^6}{36 m_{\Omega_b^*} m_V^2 m_b} + \frac{m_c^6}{36 m_{\Omega_c^*} m_V^2 m_b} \Big] \nonumber \\&
		- \frac{4}{m_{\Omega_b^*}^6} G_5^2(m_V^2)
		\Big[
		-\frac{16}{9} m_{\Omega_c^*}^2 m_b^3 + \frac{16 m_{\Omega_c^*}^4 m_b^3}{9 m_V^2} 
		+ \frac{8}{9} m_{\Omega_c^*}^2 m_b^2 m_c - \frac{8 m_{\Omega_c^*}^4 m_b^2 m_c}{9 m_V^2} 
		- \frac{16 m_{\Omega_c^*}^2 m_b^3 m_c}{9 m_V^2} \nonumber \\&
		+ \frac{8}{9} m_{\Omega_c^*}^2 m_b m_c^2 - \frac{8 m_{\Omega_c^*}^4 m_b m_c^2}{9 m_V^2} 
		+ \frac{8 m_{\Omega_c^*}^2 m_b^2 m_c^2}{9 m_V^2} + \frac{4 m_b^3 m_c^2}{9 m_V^2} 
		- \frac{4}{9} m_{\Omega_c^*}^2 m_c^3 + \frac{4 m_{\Omega_c^*}^4 m_c^3}{9 m_V^2} \nonumber \\&
		+ \frac{8 m_{\Omega_c^*}^2 m_b m_c^3}{9 m_V^2} - \frac{2 m_b^2 m_c^3}{9 m_V^2} 
		- \frac{m_{\Omega_c^*} m_c^4}{9 m_{\Omega_b^*}} - \frac{4 m_{\Omega_c^*}^2 m_c^4}{9 m_V^2} 
		+ \frac{m_{\Omega_c^*}^3 m_c^4}{9 m_{\Omega_b^*} m_V^2} - \frac{2 m_b m_c^4}{9 m_V^2} \nonumber \\&
		+ \frac{m_c^5}{18 m_{\Omega_b^*}^2} + \frac{m_c^5}{9 m_V^2} - \frac{m_{\Omega_c^*} m_c^5}{9 m_{\Omega_b^*} m_V^2} 
		- \frac{m_{\Omega_c^*}^2 m_c^5}{18 m_{\Omega_b^*}^2 m_V^2} + \frac{m_c^6}{18 m_{\Omega_b^*}^2 m_V^2} 
		+ \frac{m_c^6}{36 m_V^2 m_b} - \frac{m_c^7}{72 m_V^2 m_b^2}
		\Big] \nonumber \\& + \frac{4}{m_{\Omega_b^*}^{6}} F_5^2(m_V^2)
		\Big[
		-\frac{16}{9} m_{\Omega_c^*}^2 m_b^3 + \frac{16 m_{\Omega_c^*}^4 m_b^3}{9 m_V^2} 
		- \frac{8}{9} m_{\Omega_c^*}^2 m_b^2 m_c + \frac{8 m_{\Omega_c^*}^4 m_b^2 m_c}{9 m_V^2} 
		- \frac{16 m_{\Omega_c^*}^2 m_b^3 m_c}{9 m_V^2} \nonumber \\&
		+ \frac{8}{9} m_{\Omega_c^*}^2 m_b m_c^2 - \frac{8 m_{\Omega_c^*}^4 m_b m_c^2}{9 m_V^2} 
		- \frac{8 m_{\Omega_c^*}^2 m_b^2 m_c^2}{9 m_V^2} + \frac{4 m_b^3 m_c^2}{9 m_V^2} 
		+ \frac{4}{9} m_{\Omega_c^*}^2 m_c^3 - \frac{4 m_{\Omega_c^*}^4 m_c^3}{9 m_V^2} \nonumber \\&
		+ \frac{8 m_{\Omega_c^*}^2 m_b m_c^3}{9 m_V^2} + \frac{2 m_b^2 m_c^3}{9 m_V^2} 
		- \frac{m_{\Omega_c^*} m_c^4}{9 m_{\Omega_b^*}} + \frac{4 m_{\Omega_c^*}^2 m_c^4}{9 m_V^2} 
		+ \frac{m_{\Omega_c^*}^3 m_c^4}{9 m_{\Omega_b^*} m_V^2} - \frac{2 m_b m_c^4}{9 m_V^2} \nonumber \\&
		- \frac{m_c^5}{18 m_{\Omega_b^*}^2} - \frac{m_c^5}{9 m_V^2} - \frac{m_{\Omega_c^*} m_c^5}{9 m_{\Omega_b^*} m_V^2} 
		+ \frac{m_{\Omega_c^*}^2 m_c^5}{18 m_{\Omega_b^*}^2 m_V^2} - \frac{m_c^6}{18 m_{\Omega_b^*}^2 m_V^2} 
		+ \frac{m_c^6}{36 m_V^2 m_b} + \frac{m_c^7}{72 m_V^2 m_b^2} \Big] \Big\} \,,
\end{align}
where the quantities $m_+$, $m_-$, $m_a$, $m_b$, and $m_c$ are defined as follows:
\begin{align}
	&m_+ = m_{\Omega_{b}^{*}}+m_{\Omega_{c}^{*}} \,, \ \ \   m_- = m_{\Omega_{b}^{*}}-m_{\Omega_{c}^{*}} \,, \ \ \
	m_a = m_{\Omega_{b}^{*}}^2+m_{\Omega_{c}^{*}}^2-m_P^2  \,, \nonumber \\&  m_b = m_{\Omega_{b}^{*}}m_{\Omega_{c}^{*}}  \,, \ \ \ m_c = m_{\Omega_{b}^{*}}^2+m_{\Omega_{c}^{*}}^2-m_V^2 \,.
\end{align}
In the derivation of Eqs.~(\ref{eq:SquAmPseM}) and~(\ref{eq:SquAmVecM}), the following relations have been employed~\cite{Amiri:2025gcf,Duan:2024zjv}:
\begin{align} \label{RSSum}
	\sum_{s}u_{\beta}^{\Omega_{b}^{*}}(p,s)\ \bar{u}_{\nu}^{\Omega_{b}^{*}}(p,s)  =& -\big(\slashed{p}+ m_{\Omega_{b}^{*}}\big)\ \Big[g_{\beta\nu}-\frac{1}{3}\gamma_{\beta}\gamma_{\nu}-\frac{2}{3}\frac{p_{\beta}p_{\nu}}{m_{\Omega_{b}^{*}}^{2}}+\frac{1}{3}\frac{p_{\beta}\gamma_{\nu}-p_{\nu}\gamma_{\beta}}{m_{\Omega_{b}^{*}}}\Big]  \,, \nonumber \\
	\sum_{s'}u_{\rho}^{\Omega_{c}^{*}}(p',s')\ \bar{u}_{\alpha}^{\Omega_{c}^{*}}(p',s')  =& -\big(\slashed{p'}+ m_{\Omega_{c}^{*}}\big)\ \Big[g_{\rho\alpha}-\frac{1}{3}\gamma_{\rho}\gamma_{\alpha}-\frac{2}{3}\frac{p'_{\rho}p'_{\alpha}}{m_{\Omega_{c}^{*}}^{2}}+\frac{1}{3}\frac{p'_{\rho}\gamma_{\alpha}-p'_{\alpha}\gamma_{\rho}}{m_{\Omega_{c}^{*}}}\Big] \,, \nonumber \\
	\sum_\lambda \epsilon_\lambda^{*\mu}(q) \epsilon_\lambda^{\nu}(q) =& -g^{\mu\nu}+\frac{q^{\mu}q^{\nu}}{m_V^2} \,.
\end{align}

\bibliographystyle{JHEP}
\bibliography{bibliography}

\end{document}